\documentclass[a4paper,11pt]{article}
\usepackage{jheppub}
\usepackage[T1]{fontenc}
\usepackage[utf8]{inputenc}
\usepackage{amsmath}
\usepackage{amsfonts}
\usepackage{amssymb}
\usepackage{graphicx}
\usepackage{relsize}
\usepackage{pstricks}
\usepackage{color}
\usepackage{tikz}
\usepackage{bm}
\usepackage{booktabs}
\usepackage{csquotes}
\usepackage{blindtext}
\usepackage{imakeidx}
\indexsetup{firstpagestyle=empty}
\makeindex[columns=2,title=Alphabetical Index,intoc]

\usepackage{epsfig}             
\usepackage{graphicx}
\usepackage{tabularx}
\usepackage{orcidlink}
\usepackage{dsfont}
\usepackage[makeroom]{cancel}
\usepackage[T1]{fontenc}

\newcommand{\seq}{\begin{subequations}}
\newcommand{\sen}{\end{subequations}}
\newcommand{\eq}{\begin{eqnarray}}
\newcommand{\en}{\end{eqnarray}}
\newcommand{\la}{\langle}
\newcommand{\ra}{\rangle}

\newcommand{\Ma}{M_{a,0}}
\newcommand{\Tr}{{\rm Tr}}

\begin{document}

\title{\boldmath Probing leptophobic dark sector with \\
  a pseudoscalar portal in the NA64 experiment \\ at CERN} 

\author[a]{Sergei N.~Gninenko\,\orcidlink{0000-0001-6495-7619},}
  
\author[a]{Dmitry V.~Kirpichnikov\,\orcidlink{0000-0002-7177-077X},} 

\author[a]{Nikolai V.~Krasnikov\,\orcidlink{0000-0002-8717-6492},} 

\author[b,c]{Sergey Kuleshov\,\orcidlink{0000-0002-3065-326X},}

\author[d,b,1]{Valery E. Lyubovitskij\note{Corresponding author}
\,\orcidlink{0000-0001-7467-572X},}

\author[e,f]{Alexey S. Zhevlakov\,\orcidlink{0000-0002-7775-5917}}

\affiliation[a]{Institute for Nuclear Research of the Russian Academy 
  of Sciences, 117312 Moscow, Russia}

\affiliation[b]{Millennium Institute for Subatomic Physics at 
the High-Energy Frontier (SAPHIR) of ANID, \\
Fern\'andez Concha 700, Santiago, Chile}

\affiliation[c]{Center for Theoretical and Experimental Particle Physics,
  Facultad de Ciencias Exactas, \\ Universidad Andres Bello,
  Fernandez Concha 700, Santiago, Chile}

\affiliation[d]{Institut f\"ur Theoretische Physik,
        Universit\"at T\"ubingen, \\
	Kepler Center for Astro and Particle Physics, \\ 
	Auf der Morgenstelle 14, D-72076 T\"ubingen, Germany}

\affiliation[e]{Bogoliubov Laboratory of Theoretical Physics, JINR,
 141980 Dubna, Russia} 

\affiliation[f]{Matrosov Institute for System Dynamics and 
 Control Theory SB RAS, \\  Lermontov str., 134,
 664033, Irkutsk, Russia }

\emailAdd{sergei.gninenko@cern.ch}
\emailAdd{kirpich@ms2.inr.ac.ru}
\emailAdd{nikolai.krasnikov@cern.ch}
\emailAdd{serguei.koulechov@cern.ch}
\emailAdd{valeri.lyubovitskij@uni-tuebingen.de}
\emailAdd{zhevlakov@theor.jinr.ru}

\abstract{
We propose the possibility of discovering a light pseudoscalar 
particle $a$, axion-like particle (ALP),
interacting mainly with quarks using the electron and photon scattering
reaction chain $e + Z \rightarrow e + \gamma + Z$;
$\gamma + Z \rightarrow a + Z$ on nuclei in the NA64 experiment
at the CERN SPS. We consider the mixing of ALP with light pseudoscalar
mesons $P = \pi, \eta, \eta'$ with taking into account of the ALP mass
explicitly breaking the Peccei-Quinn symmetry. It could open invisible
channels of ALP decaying into dark fermion matter by analogy with 
$P$ mesons decays. New bounds on the coupling strengths of the ALP 
with quarks and of pseudoscalar mesons 
with dark fermions are obtained by using existing upper bounds on
invisible decay modes of $P$s  including those recently derived by NA64. 
We also study a scenario when the $a$ plays the role of a messenger in
the communication between our world and the dark sector.
New upper limits on axion couplings with quarks and its mixing
parameters with $\pi, \eta, \eta'$ mesons are established.}  


\flushbottom

\maketitle

\pagenumbering{roman}
\setcounter{page}{2}
\clearpage
\pagenumbering{arabic}
\setcounter{page}{1}

\section{Introduction}
\label{sec:introduction}

Pseudoscalar mesons  play important role in contemporary particle
physics. In particular, due to the nontrivial
structure of quantum chromodynamics (QCD) vacuum
eight light pseudoscalar mesons ($\pi$, $K$, $\eta$) appear as
Goldstone massless particles associated with spontaneous breaking
of chiral symmetry. The source of explicit breaking of chiral symmetry
is the current quark masses $m_q$. In particular, the masses of
$\pi$, $K$, and $\eta$ at the leading order of chiral expansion are
proportional to $m_q$ and the quark condensate, the parameter of
spontaneous breaking of chiral symmetry. The dynamics of light
pseudoscalars is well and consistently described by the chiral
perturbation theory (ChPT)~\cite{Weinberg:1978kz,Gasser:1983yg,Gasser:1984gg}.

The QCD vacuum is also responsible for a CP violation in the QCD Lagrangian,
producing the so-called $\theta$-term. The $\theta$-term
is induced by the instanton effects~\cite{Polyakov:1975rs}-\cite{Coleman:1979} 
and related to such important physical phenomena as the nonvanishing electric
dipole moment of the neutron and rare two-pion decays of the $\eta$ and
$\eta'$ mesons~\cite{Baluni:1978rf}-\cite{Faessler:2006a}. 
To resolve the strong CP-violation in QCD 
the Peccei-Quinn (PQ) mechanism has been proposed~\cite{Peccei:1977hh,Peccei:1977gg}.  
This mechanism is based on the idea of existing a new type of QCD symmetry
(the PQ symmetry), which is spontaneously broken at scale much larger than
the electroweak scale to be consistent with data, astrophysical and cosmological
phenomena~\cite{Peccei:2006as,Castillo-Felisola:2015ema}. It gives rise
a new pseudoscalar particle, axion~\cite{Wilczek:1977pj,Weinberg:1977ma}.
The QCD axion mass has been be calculated using chiral algebra
leading to two equivalent results~\cite{Weinberg:1977ma,Wilczek:1977pj}
and~\cite{Shifman:1979if}, which are differed by interpretation
of the vacuum expectation of the scalar field, which is related to the
effective Fermi constant $G_F$ in~\cite{Weinberg:1977ma} and
a free parameter in~\cite{Shifman:1979if}.
Finally, restricting to the two flavor ($u$, $d$) picture
one can express the QCD axion mass
as $M_{a, {\rm QCD}}^2 = (F_\pi/f_a)^2 M_\pi^2 m_um_d/(m_u+m_d)^2$,
where $f_a$ is the axion decay constant,
$F_\pi = 0.0924$ GeV and $M_\pi$ are the pion decay constant and mass,
$m_u$ and $m_d$ are the current masses of the $u$ and $d$ quarks,
respectively. 
The vacuum expectation value of the axion can be fixed to cancel the
theta parameter and solve strong CP-violation problem. 
During past decades the QCD axion and its extension to the sector of so-called
axion-like particles (ALPs) were extensively used for possible resolutions of
existing puzzles in astro-particle and particle physics, and cosmology starting
from the strong CP-violation problem, including rare processes,
and considerations of them as a portal to Dark Matter (DM).

The mass of the ninth pseudoscalar $\eta'$ is much larger than their
eight partners ($\pi$, $K$, $\eta$) and it can be included in QCD using
large $N_c$ arguments indicating that QCD at large $N_c$ is
reasonable approximation to the real world of strong
interactions~\cite{tHooft:1973alw}-\cite{Kaiser:2000}. 
Using $1/N_c$ expansion one can solve $U_A(1)$ problem without instantons and
relate it to existence of the ninth Goldstone boson,
$\eta'$ meson~\cite{Witten:1979vv,Veneziano:1979ec}. 
In particular~\cite{Witten:1979vv}, it was argued that at large $N_c$ the axial
$U_A(1)$ symmetry is exact, the corresponding current is conserved,
and $\eta'$ meson manifests itself as a massless $U_A(1)$ Goldstone. 
The $U_A(1)$ symmetry is violated by the gluon anomaly, which together with the
$\eta'$ meson mass $M_{\eta'}$ scales as $1/N_c$ and vanishes at $N_c \to \infty$.
Moreover, at large $N_c$ limit the soft $\eta'$ theorems take place, which have
analogy with the soft pion theorems. Later, in Ref.~\cite{Veneziano:1979ec},
it was shown that ideas of Ref.~\cite{Witten:1979vv} are consistent with
expected $\theta$ dependences and anomalous Ward identities.

In order to involve the $\eta'$ state in the low-energy chiral Lagrangians,
the Large $N_c$ ChPT has been developed
in Refs.~\cite{DiVecchia:1979pzw,Nath:1981,Leutwyler:1997yr,Leutwyler:1996np,%
Kaiser:1998ds,Kaiser:2000,Witten:1980sp,DiVecchia:1980yfw,Rosenzweig:1980}.   
In particular, it was shown that the $\eta'$ meson can be included in the
ChPT Lagrangian doing replacement of the $SU(3)$ chiral field by one
representing an element $U(x)$ of the $U(3)$ group, whose phase $\psi$ defined
as ${\rm det}U(x) = e^{i \psi(x)}$ and is identified with singlet field $\eta^0$.
Effective Lagrangian is constructed in terms of the chiral field $U(x)$ and
expanded in powers of $1/N_c$, powers of the momenta, and current quark masses.
Physical states of $\eta$ and $\eta'$ meson appear as result of the
$\eta^0-\eta^8$ mixing. Large mass of the $\eta'$ meson is explained by
the sizable contribution of the $U_A(1)$ anomaly $m_0^2$ or the topological
susceptibility $\tau$ of the purely gluonic theory. The quantity $\tau = m_0^2$
is counted in the $1/N_c$ expansion as ${\cal O}(N_c^0)$.
In Ref.~\cite{Leutwyler:1996np} the $\pi^0-\eta-\eta'$ mixing has been studied.
The mixing of three pseudoscalars was represented in terms of three angles:
$\epsilon = \theta_{\pi\eta}$, $\epsilon' = \theta_{\pi\eta'}$,
and $\theta_{\eta\eta'}$. 
It was shown that the isospin breaking angles $\epsilon$ and $\epsilon'$ are
proportional to the quark mass ratio $(m_d - m_u)/(m_s - \hat{m})$,
where $\hat{m} = (m_u+m_d)/2$,
and the combinations of the $\cos\theta_{\eta\eta'}$ and $\sin\theta_{\eta\eta'}$. 
Another important finding of the Large $N_c$
QCD is the mixing of the leptonic decay constants of the
$\eta$ and $\eta'$ mesons~\cite{Kaiser:1998ds}, which are parametrized by two 
mixing angles $\theta_0$ (singlet) and $\theta_8$ (octet) specifying
the $\eta$ and $\eta'$ projections of the states $A_\mu^0 |0\ra$ and
$A_\mu^8 |0\ra$, where $A_\mu^0$ and $A_\mu^8$ are the $U(3)$ singlet and octet
axial-vector quark currents. Later importance of such a mixing of the
leptonic decay constants of the $\eta$ and $\eta'$ mesons was investigated
and proved in Refs.~\cite{Feldmann:1998vh,Feldmann:1999,Feldmann:1998,Feldmann:1999uf}
in the analysis of the two-photon decay rates, $P\gamma\gamma$ and $VP\gamma$ transition
form factors. 

In this work we study the possibility of discovering ALP interacting mainly
with quarks and light dark matter with the NA64 experiment at the CERN SPS.
It is a continuation of a series of our papers on the searching for
new physics by using  the charge exchange reactions
at NA64~\cite{Gninenko:2023rbf,Zhevlakov:2023wel}. 
Namely, we consider the electron and photon scattering reaction chain  
\begin{equation}
e + Z\rightarrow e + \gamma + Z; \gamma + Z \rightarrow a + Z
\label{eq:chain}
\end{equation}
on nuclei as a source of $a$'s, and obtain new bounds
on the $a$ coupling strength with quarks and new limits
on invisible pseudoscalar meson $P$ ($P = \pi^0, ~\eta, ~\eta'$)
decays using the corresponding  constraints on~\eqref{eq:chain}
from the NA64~\cite{Andreev:2024tmh}. 
In particular, the novel technique of the NA64h experiment 
uses the charge-exchange neutral mesons. 
The $\eta, \eta' \to invisible events$ would exhibit themselves 
via a striking signature - the complete disappearance of the incoming
beam energy in the detector. This allowed to the NA64h experiment to set
a stringent limit on the branching ratio
${\rm Br}(\eta' \to invisible) < 2.1 \times 10^{-4}$~\cite{Andreev:2024tmh} 
improving the current bound by a factor of $\simeq 3$.
It was also set a limit on
${\rm Br}(\eta \to invisible) < 1.1 \times 10^{-4}$
comparable with the existing one.
These results demonstrated the great potential of the NA64h experiment and
provided clear guidance on how to enhance and extend the sensitivity for
dark sector physics from future searches for invisible neutral meson decays.  
Besides, the $a-P$  mixing is interesting due to
future experimental facilities ($\eta$ and $\eta^\prime$ factories)
in USA in FermiLab (REDTOP project)~\cite{REDTOP:2022slw} and
in China at Huizhou (HIAF project)~\cite{Chen:2024wad}.
Both projects aim to precision studies of $\eta$ and $\eta^\prime$
meson rare decays including search and test of new physics. 
Also we study the possibility that the $a$ plays the role
of a messenger in the communication between our world and dark sector.

In this vein, we take into account the $a-P$  mixing, previously  discussed 
in the literature in context of QCD axion, see, e.g., 
Refs.~\cite{Georgi:1986df},\cite{Krauss:1986bq}-\cite{Aghaie:2024jkj}
based on the chiral Lagrangian. For the first time, this Lagrangian was derived
in Ref.~\cite{Georgi:1986df} as an extension of
the ChPT Lagrangian~\cite{Weinberg:1978kz,Gasser:1983yg,Gasser:1984gg,Leutwyler:1997yr,Kaiser:2000},  
where the derived $a-P$ mixing results in the 
producing of two types of the mixing term: (1) mass mixing 
via replacement the quark mass matrix ${\cal M}={\rm diag}(m_u,m_d,m_s)$
by the axion-dependent matrix
${\cal M}(a) = {\rm diag}(m_u e^{i a \frac{Q_u}{f_a}},
                          m_d e^{i a \frac{Q_d}{f_a}},
                          m_s e^{i a \frac{Q_s}{f_a}})$,
where $Q_q$ with $q=u,d,s$ are the PQ charges of the quarks;
(2) kinetic mixing after the axial transformation removing the anomaly
coupling of the axion field with gluons and producing the 
derivative coupling of the axion to axial-vector currents composed
of the chiral field matrix $U(x)$.
In order to deal with physical states of axion and pseudoscalar fields one
should perform diagonalization of the mixing term of the chiral Lagrangian.
In this vein, one can use different approximations and assumptions.
E.g., in Ref.~\cite{Georgi:1986df} it was suggested to postulate the PQ charge
matrix of quark in the form
${\cal Q}_{q} = {\rm diag}(Q_u,Q_d,Q_s) = {\cal M}^{-1}/\Tr({\cal M}^{-1})$.
Such ansatz for the  $Q_{\rm PQ}$ allows to
eliminate the mass mixing term between
the axion and the pseudoscalar mesons. Later on, the formalism proposed in
Ref.~\cite{Georgi:1986df} was extensively used for physical applications
involving axion/ALP and pseudoscalar mesons
($\pi^0$, $\eta$, $\eta'$)~\cite{Krauss:1986bq}-\cite{Aghaie:2024jkj}
considering various limiting cases, e.g., restricting to the schemes:
(1) involving the axion and one (or two) pseudoscalars;
(2) neglecting kinetic or mass mixing of the axion with some of
the pseudosclars; (3) neglecting mass mixing between some of
the pseudoscalars. 
The most complete and detailed mixing scheme involving the axion and all three 
pseudoscalars ($\pi^0$, $\eta$, $\eta'$) has been considered for the first time
in Ref.~\cite{Choi:1986zw} and later in Ref.~\cite{Aloni:2018vki}. Note, in
Ref.~\cite{Choi:1986zw} the framework was limited to consideration of the QCD
axion by neglecting the PQ-breaking contribution to its mass $\Ma$.  
In Ref.~\cite{Aloni:2018vki} the $\eta-\eta'$ mixing was restricted to the
ideal mixing with $\cos\theta_{\eta\eta'}^I = 1/\sqrt{3}$. In our paper we will
follow Ref.~\cite{Alves:2024dpa} and consider nonderivative coupling
of the axion to the pseudoscalars via replacement the quark mass matrix
${\cal M}$ by the axion-dependent matrix ${\cal M}(a)$, also we include 
the contribution of the gluon and heavy quarks $c, b, t$ by integrating over
these degrees of freedom, the contribution of the axial $U_A(1)$ anomaly 
to the $\eta^0$ mass, go beyond the limit of the ideal $\eta-\eta'$ mixing 
and take into account $\Ma \neq 0$. 
Note that taking into account of the finite ALP mass  $\Ma \neq 0$
has been done before in Refs.~\cite{Aloni:2018vki,Alves:2024dpa} 
applying specific approximations. 
In particular, in Ref.~\cite{Aloni:2018vki} the ideal mixing
of the $\eta-\eta'$ system was considered and a mixing of the $\pi^0$
with $\eta$ and $\eta'$ was neglected. In Ref.~\cite{Alves:2024dpa}  
the ALP-pseudoscalar mesons mixing angles were derived
by neglecting $\pi-\eta$ and $\pi-\eta'$ mixing
(i.e., making the approximation of $\theta_{\eta\pi^{(')}} \to 0$). 
In our consideration we will derive the ALP-pseudoscalar mesons 
mixing angles at $\Ma \neq 0$ with additional approximations.  
Therefore, our paper is well-motivated by recent progress 
in experimental study of invisible decays of 
pseudoscalar mesons into dark matter~\cite{Andreev:2024tmh}, 
which open a possibility for searching for invisible
decays for massive ALP. Note, invisible decays of massive ALP  
are not kinematically forbidden in comparison with very light 
QCD axion. On the other hand, NA64h setup of the 
NA64 Collaboration~\cite{Andreev:2024tmh} gives the  
opportunity for study of invisible decays of ALPs.  
  
Our paper is organized as follows. In Sec.~\ref{sec:effLarg}
we present the derivation of the effective Lagrangian describing
ALP, light pseudoscalar mesons and their mixing.
Diagonalizing the mixing terms we get the set of physical states
of ALP, $\pi^0$, $\eta$, and $\eta'$.
In Sec.~\ref{DM_field} we discuss the inclusion of light Dark Matter
fermions and its coupling to ALP and pseudoscalar fields.
In Sec.~\ref{ALP_2photons} we derive the anomalous couplings 
of ALP and pseudoscalar mesons with two photons. 
In Sec.~\ref{sec:results} we consider the application of our formalism
to deriving constraints on the mixing parameters of ALP with pseudoscalar
mesons and combinations of the PQ charges using invisible
decays of $\pi^0$, $\eta$, and $\eta'$ into a pair light DM particles,
cross sections of the scattered DM particles into quarks and hadrons. 
In Sec.~\ref{sec:summary} we present the summary of
our paper. Some calculation details including
the diagonalization of mass term of ALP and pseudoscalar mesons
are collected in the Appendix.

\section{ALP-Pseudoscalar Mesons Effective Lagrangian}
\label{sec:effLarg}

Study of the processes involving ALP and light pseudoscalar
mesons  $P $ will be perform
using the effective Lagrangian, which was derived in Ref.~\cite{Georgi:1986df}
after mapping of the QCD-level quark-gluon Lagrangian into meson ChPT
Lagrangian~\cite{Weinberg:1978kz,Gasser:1983yg,Gasser:1984gg,Leutwyler:1997yr,Kaiser:2000}.   
As we stressed in Introduction we will follow ideas
of Ref.~\cite{Georgi:1986df}, where this mapping was proposed and 
Refs.~\cite{Krauss:1986bq}-\cite{Aghaie:2024jkj}, which further developed
the formalism presented in~\cite{Georgi:1986df}. In our consideration
we include all terms of the resulting Lagrangian producing
mixing between four states: ALP, $\pi^0$, $\eta$, and $\eta'$.
  
The ALP-pseudoscalar meson leading-order ChPT Lagrangian based on
the non-derivative coupling of the ALP to the light pseudoscalar mesons 
reads~\cite{Alves:2024dpa}:  
\eq\label{ALP_P_Lagrangian} 
{\cal L}_{\rm ALP+P} &=&
\frac{1}{2} \, \Big[\partial_\mu a^0 \partial^\mu a^0
  -  \Ma^2 \, (a^0)^2\Big]   
  + \frac{F^2}{4}  \Tr\Big[D_{\mu}U^{\dagger} \, D^{\mu}U^{\dagger}\Big]
\nonumber\\
&+&  \frac{F^2 B}{2} \, \Tr\Big[{\cal M}(a^0) \, U + {\cal M}^\dagger(a^0) \,
U^{\dagger}\Big] - \frac{\tau}{2}
\Big[- i \log \, {\rm det}U - \frac{Q_{GQ}}{f_{a}} \, a^0\Big]^2 \,, 
\en
where $a$ is the ALP field,
$U$ is the chiral field in the nonlinear exponential representation collecting
nine pseudoscalars ($\pi$, $K$, $\eta^8$, $\eta^0$):
\eq
U = \exp\biggl[\frac{i P}{F}\biggr]\,, \qquad 
P = \sqrt{2} 
\begin{pmatrix}
  \frac{\pi^3}{\sqrt{2}}+\frac{\eta^8}{\sqrt{6}}+\frac{\eta^0}{\sqrt{3}}
   &\pi^+ & K^+\\[2mm]
  \pi^- & -\frac{\pi^3}{\sqrt{2}}+\frac{\eta^8}{\sqrt{6}}
  + \frac{\eta^0}{\sqrt{3}}
  & K^0\\[2mm]
  K^-   & \overline{K}^{0} 
  & -\frac{2 \eta^8}{\sqrt{6}} + \frac{\eta^0}{\sqrt{3}} 
\end{pmatrix}\,, 
\label{U_field}
\en
and $F$ is the leptonic decay constant of pseudoscalar mesons
in the chiral limit, 
$D_{\mu}$ is the covariant derivative acting on chiral field and including
external vector $v_\mu$ and axial-vector $a_\mu$ fields, which is defined as 
\eq
D_\mu U = \partial_\mu U - i (v_\mu + a_\mu) U + i U (v_\mu - a_\mu) \,. 
\en
Diagonal matrix 
\eq
{\cal M}(a^0) = {\rm diag}\Big(e^{iQ_{u}a^0/f_{a}} \, m_u, e^{iQ_{d}a^0/f_{a}}
\, m_d, e^{iQ_{s}a^0/f_{a}} \, m_s\Big)
\en   
collects the ALP-dependent quark masses, where
$f_a$ is the ALP decay constant 
and $Q_q$ is the PQ charge of the quark of flavor $q=u,d,s$.  
Let us discuss the meaning of the terms in the effective
Lagrangian~(\ref{ALP_P_Lagrangian}).
The first term is the free Lagrangian of the ALP
with the mass $\Ma$. 
Note, in the limit $\Ma = 0$ the ALP is identified with the QCD axion,
which predominantly decays into two photons, while other processes are
kinematically forbidden due to very small mass of the QCD axion. 
The use of $\Ma \neq 0$ allows us to go beyond this picture
and propose the processes, where the ALP decays into Dark Matter,
e.g. dark fermions. 

The second term is the leading-order chiral- and gauge-invariant ChPT
Lagrangian~\cite{Weinberg:1978kz,Gasser:1983yg,Gasser:1984gg} 
describing the dynamics of the nonet of massless pseudoscalar mesons
($\pi$, $K$, $\eta^8$, $\eta^0$) and their couplings with external
vector and axial fields.
The third term describes the mass term of pseudoscalar mesons
and their nonderivative coupling with the ALP.
This term generates the so-called
mass mixing of the ALP with pseudoscalar mesons~\cite{Georgi:1986df}.
Finally, the fourth term is generated by the
$U_A(1)$ anomaly, where $\tau$ is the topological susceptibility of the
purely gluonic theory~\cite{Leutwyler:1997yr}-\cite{DiVecchia:1980yfw} 
and $- \frac{Q_{GQ}}{\sqrt{6}} \, \frac{F}{f_{a}}$ is the contribution of
the pure gluonic theory and heavy quarks $Q=c,b,t$~\cite{Alves:2024dpa},
which are integrated out. Here $Q_{GQ} = Q_G + Q_c + Q_b + Q_t$
is the sum of the bare PQ charge of the gluons $(Q_G)$ and
heavy quarks $(Q_c, Q_b, Q_t)$. Taking into account that
$- i \log \, {\rm det}U = (\eta^0 \sqrt{6})/F$ we finally
get~\cite{Alves:2024dpa} for the fourth term: 
\eq 
\Delta {\cal L} = -  \frac{M_0^2}{2} \,
\Big[\eta^0 - \frac{Q_{GQ}}{\sqrt{6}} \, \frac{F}{f_{a}} \, a^0\Big]^2  \,, 
\en 
where $M_0^2 = 6 \tau/F^2$ is the leading-order contribution
to the $\eta^0$ meson mass squared due to the axial anomaly.
Appearance of the ALP-dependent term follows from
Refs.~\cite{Witten:1980sp,DiVecchia:1980yfw}, where the shift 
$\log \, {\rm det}U \to \log \, {\rm det}U - \theta$ was produced
by two equivalent methods. Here, if we substitute the $\theta$ term
by the $\theta_{\rm eff}(x) = \theta + Q_{GQ} \, (F/f_{a}) \, a^0(x)$,
linear combination of the $\theta$ term
and ALP, then we get the Lagrangian $\Delta {\cal L}$,
which describes the $U(1)$ anomaly contribution to the
masses of the $\eta^0$ meson and ALP and their mixing. Note, that 
the effective field $\theta_{\rm eff}(x)$ does not contain effects of
light quarks to exclude a double-counting. Diagonalization of 
the mass term involving bare ALP field $a^0$ and 
the pseudoscalars $(\pi^3, \eta^8, \eta^0)$ leading to the
mixing of physical states $a, \pi^0, \eta, \eta'$ is discussed
in detail in Appendix~\ref{App:A}. We perform diagonalization
for $\Ma \neq 0$ and for completeness consider limit $\Ma = 0$. 
In particular, we derived expression  
for the ALP mass in terms of parameter $\Ma$~(\ref{MA2_full}) 
and set of mixing parameters ($\theta_{aP}$, $P = \pi^0, \eta, \eta'$)
of ALP with pseudoscalar
mesons~(\ref{theta_aP_full})-(\ref{theta_aP_full2}), (\ref{theta_aP_exp}).

\begin{figure}
\begin{center}
\epsfig{figure=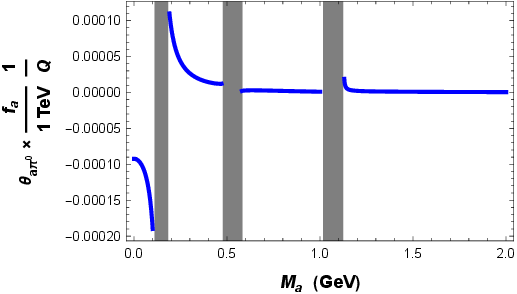,scale=1}

\vspace*{.75cm}
\epsfig{figure=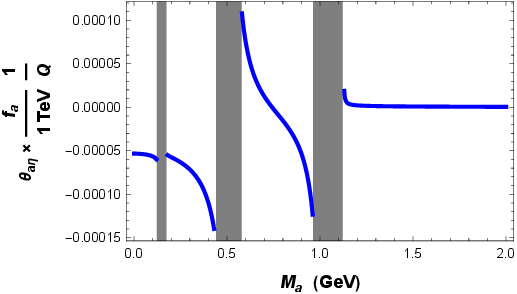,scale=1}

\vspace*{.75cm}
\epsfig{figure=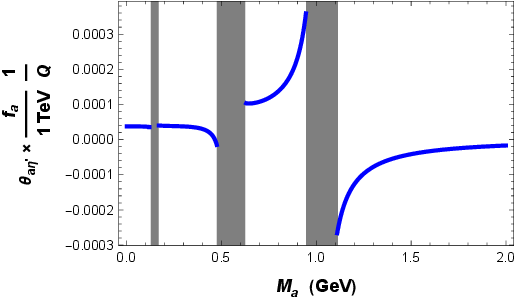,scale=1}

\end{center}
\noindent
\caption{Predictions for the ALP-PS mesons mixing
  parameters as functions of the ALP mass~$M_a$.
  Shaded regions correspond to the poles of the mixing
  angles. 
    \label{fig1}}
\end{figure}

In Fig.~\ref{fig1}  
we display the results for
the partial contributions to the mixing parameters
$\theta_{a\pi^0}$, $\theta_{a\eta}$, and
$\theta_{a\eta'}$ as functions of the ALP mass $M_a$.   
In the plots we factorize out
the factor $\frac{f_a}{\rm 1 TeV} \frac{1}{Q}$, 
where $f_a$ is given in the units of TeV and
$Q$ is averaged PQ charge, i.e. 
for specific values of PQs we use: $Q_u = - Q_d = - Q_s = Q_{GQ} = Q$. 
Also we shade out the regions where the ALP mass appears to be degenerate
with the $\pi^0$, $\eta$, and $\eta'$ masses.
ALP-pseudoscalar mixing parameters scale at large $\Ma$ at least
as $1/\Ma^2$.

\section{Inclusion of Dark Fermion Field}
\label{DM_field}

In this section we discuss inclusion of the Dark fermions 
in our effective Lagrangian~(\ref{ALP_P_Lagrangian}).
Such important extension of the effective Lagrangian is relevant
for study of invisible decay modes of ALP. As we stressed before,
these decay modes are possible for the case of massive ALP 
with taking into account the mass $\Ma$, 
the leading perturbative contribution to the ALP mass due explicit
breaking of the PQ symmetry. 

We assume that Dark Matter (DM) is described by Dirac fermion $\chi$
with mass $M_\chi$. 
Lagrangian involving field $\chi$ has two main terms:
(1)~free Lagrangian $L_{\chi, 0}$ and
(2)~interaction Lagrangian ${\cal L}_{a \chi \bar{\chi}}$
of $\chi$ with ALP:
\eq
{\cal L}_{\chi} &=& {\cal L}_{\chi, 0} + {\cal L}_{a \chi \bar{\chi}}
\,, \nonumber\\
  {\cal L}_{\chi, 0} &=& \bar{\chi}
\ (i \not\!\partial - M_{\chi}) \ \chi
\,, \nonumber\\
  {\cal L}_{a \chi \bar{\chi}} &=&
  \sum\limits_{\Gamma = s, p}
             {\cal L}_{a \chi \bar{\chi}}^{\Gamma}\,,
             \quad
             {\cal L}_{a \chi \bar{\chi}}^{\Gamma} = 
             g_{\Gamma} \, a \, J_\Gamma \,, 
\label{chia}
\en
where 
$J_s = \bar{\chi}  \, \chi$ and 
$J_p = \bar{\chi}  \, i \gamma_5  \, \chi$ are the
scalar and pseudoscalar currents composed of DM fermions,
$g_s$ and $g_p$ are the coupling constants,
corresponding to the $J_s$ and $J_p$ currents.  

Using interaction Lagrangian~${\cal L}_{a \chi \bar{\chi}}$
for $M_a  > 2 M_{\chi}$ we can calculate the decay width
$\Gamma(a \rightarrow \chi \bar{\chi})$,
which is given by the formula 
\eq\label{achidecay} 
\Gamma(a  \rightarrow \chi \bar{\chi})
= \frac{M_a}{8\pi} \, \beta_{a\chi}
\, \Big[g^2_{p} + \beta_{a\chi}^2 \, g^2_{s}\Big] \,, \qquad 
\beta_{H_1H_2} = \Big(1  - \frac{4 M^2_{H_2}}{M^2_{H_1}}\Big)^{1/2} \,.
\en 
Diagonalization of the mass term in the effective
Lagrangian~(\ref{ALP_P_Lagrangian}) induces the mixing
of the ALP with neutral pseudoscalar fields of $\pi^0, \eta, \eta'$
in the couplings with other fields,  including fields of 
DM fermions. In particular, the shift of the ALP field 
\eq\label{a_pi_eta_etap_mixing}
a \to a
  - \bar\theta_{a\pi^0} \, \pi^0
  - \bar\theta_{a\eta} \, \eta
  - \bar\theta_{a\eta'} \, \eta'\,,
\en
where $\bar\theta_{a\pi^0}$ are the linear combinations of the mixing
parameters $\theta_{aP}$ of ALP $a$ with pseudoscalar mesons $P$:   
\eq
\bar\theta_{a\pi^0} &=& \theta_{a\pi^0}  \,,
\nonumber \\
\bar\theta_{a\eta} &=& \theta_{a\eta}  \, \cos\theta_{\eta\eta'}
- \theta_{a\eta'} \, \sin\theta_{\eta\eta'}  \,,
\nonumber \\
\bar\theta_{a\eta'} &=& \theta_{a\eta'} \, \cos\theta_{\eta\eta'}
      + \theta_{a\eta}  \, \sin\theta_{\eta\eta'} \,. 
\en  
The mixing parameters $\theta_{aP}$ and $\eta-\eta'$ mixing
angle are specified in Appendix~\ref{App:A}, generates the coupling
of the pseudoscalar mesons with DM fermions:
\eq\label{Lagr_Pchi}
{\cal L}_{P \chi \bar\chi} = - (g_{s} \, J_s + g_{p} \, J_p) \,
\sum\limits_{P = \pi^0, \eta, \eta'} \, \bar\theta_{aP} \, P  \,.
\en  
Therefore, Lagrangian~(\ref{Lagr_Pchi}) gives the opportunity of
invisible decays of the $P = \pi^0, \eta, \eta'$ 
into light dark matter particles $\chi \bar{\chi}$.
The corresponding decay widths are defined by the formula 
\eq 
\Gamma(P \rightarrow \chi \bar\chi) &=&
\frac{M_P}{8 \pi} \, \bar\theta^{\, 2}_{aP} \, \beta_{P\chi} \,
\Big[g_p^2 + \beta_{P\chi}^2 \, g_s^2\Big]   \,. 
\label{P2chi}
\en

\section{ALP-Two Photon Coupling}
\label{ALP_2photons}

The anomalous coupling of the pseudoscalar $P$
with two photons is defined by the effective Lagrangian 
\eq 
{\cal L}_{P\gamma\gamma} = \frac{e^2}{4} \, G_{P\gamma\gamma}
\, P \, F_{\mu \nu} \tilde{F}^{\mu\nu} \,, 
\label{P2gamma}
\en
where $F_{\mu\nu} = \partial_\mu A_\nu - \partial_\nu A_\mu$ is the stress tensor
of the electromagnetic field and 
$\tilde{F}^{\mu\nu} = \frac{1}{2} \epsilon^{\mu\nu\alpha\beta}F_{\alpha\beta}$
is its dual. 
By analogy the ALP coupling with two photons has been derived
in Ref.~\cite{Georgi:1986df}: 
\eq 
{\cal L}_{a\gamma\gamma} = \frac{e^2}{4} \, G_{a\gamma\gamma}
\, a \, F_{\mu \nu} \tilde{F}^{\mu\nu} \,. 
\label{a2gamma}
\en
Note in our definition of the couplings of ALPs and pseudoscalar
mesons with two photons we subtract the factor $e^2$ to have the
expression for the $G_{\pi^0\gamma\gamma}$ coupling consistent with
low-energy theorem $G_{\pi^0\gamma\gamma} = 1/(4 \pi^2 F)$, while
the other definition the $g_{a\gamma\gamma}$ with the factor $e^2$ hidden
in the $a(P)\gamma\gamma$ coupling is also used in the literature,
see, e.g., Ref.~\cite{NA64:2020qwq}. Therefore, two couplings
are simply related as $e^2  G_{a\gamma\gamma} =  - g_{a\gamma\gamma}$. 
Using prediction $g_{a\gamma\gamma} < 2 \times 10^{-4}$ GeV$^{-1}$
of the NA64 experiment~\cite{NA64:2020qwq} for upper limit of the
$g_{a\gamma\gamma}$ coupling,  we derive the upper limit
for the $G_{a\gamma\gamma}$ coupling:
$|G_{a\gamma\gamma}| = 2.2 \times 10^{-3}$ GeV$^{-1}$
for masses $M_a \lesssim 55$ MeV.   

Then phenomenology of the $a\gamma\gamma$ coupling was discussed
in detail in literature
(see, e.g., Refs.~\cite{Bauer:2020jbp,Aghaie:2024jkj}).
In our consideration for the first time we include three contributions to
the $a\gamma\gamma$ coupling:
(1) direct bare $a\gamma\gamma$ coupling
$G_{a\gamma\gamma,1} = G_{a\gamma\gamma}^{0}$, 
(2) the coupling $G_{a\gamma\gamma,2}$ induced by mixing with
light pseudoscalar mesons, which encodes the contribution
of light $(u,d,s)$ quarks,
(3) the coupling $G_{a\gamma\gamma,3}$ encoding
the contribution of heavy $(c,b,t)$ quarks. Therefore, the total
contribution to the $a\gamma\gamma$ couplings is
\eq
G_{a\gamma\gamma} = \sum\limits_{i=1}^3 \, G_{a\gamma\gamma,i} \,. 
\en  

The coupling $G_{a\gamma\gamma,2}$ is generated by the shifts of
the pseudoscalar fields as result of diagonalization of
the ALP-P mass term: 
\eq
\pi^0 &\to&
  \pi^0
- \theta_{\pi^0\eta}  \, \eta
- \theta_{\pi^0\eta'} \, \eta'
+ \theta_{a\pi^0}   \, a \,, 
\nonumber\\
\eta &\to&
  \cos\theta_{\eta\eta'} \, \eta
+ \sin\theta_{\eta\eta'} \, \eta' 
+ \Big(\theta_{\pi^0\eta}  \, \cos\theta_{\eta\eta'}
     + \theta_{\pi^0\eta'} \, \sin\theta_{\eta\eta'}\Big) \, \pi^0
+ \theta_{a\eta} \, a \,, 
\nonumber\\
\eta' &\to&
  \cos\theta_{\eta\eta'} \, \eta'
- \sin\theta_{\eta\eta'} \, \eta 
+ \Big(\theta_{\pi^0\eta'} \, \cos\theta_{\eta\eta'}
     - \theta_{\pi^0\eta}  \, \sin\theta_{\eta\eta'}\Big) \, \pi^0
+ \theta_{a\eta'} \, a \,.
\en
As result of such shifts we generate the second contribution to
the $a\gamma\gamma$ coupling 
\eq 
G_{a\gamma\gamma, 2} = 
\sum\limits_{P = \pi^0, \eta, \eta'} \, \theta_{aP} \, g_{P\gamma\gamma} \,.
\en
The coupling $G_{a\gamma\gamma,3}$ due to heavy quarks 
can be derived following recent paper~\cite{Aghaie:2024jkj}: 
\eq
G_{a\gamma\gamma, 3} = \sum\limits_{Q = c, b, t} \, C_Q \, e_Q^2 \,
\frac{M_a^2}{4 m_Q^2 f_a}\,,
\en
where $e_Q$ and $m_Q$ are the electric charge and mass of heavy quark.

Finally, the full expression for the $g_{a\gamma\gamma}$ coupling is
\eq\label{a2gfull} 
G_{a\gamma\gamma} = G_{a\gamma\gamma}^{0}
+ \sum\limits_{P = \pi^0, \eta, \eta'} \, \theta_{aP} \, G_{P\gamma\gamma} 
+ \sum\limits_{Q = c, b, t} \, C_Q \, e_Q^2 \, \frac{M_a^2}{4 m_Q^2 f_a}\,,
\en  
Based on the effective Lagrangians~(\ref{P2gamma}) and~(\ref{a2gamma}) a
the decay width $P  \rightarrow \gamma \gamma$ is given 
by the formula
\eq\label{decaygamma} 
\Gamma(H \rightarrow \gamma\gamma)
= \frac{\pi \, \alpha^2}{4} \, G^2_{H\gamma\gamma}
\, M_H^3 \,, \qquad H = a, P \,. 
\en
We fix the two-photon couplings of the pseudoscalar mesons
$\pi^0$, $\eta$, and $\eta'$
using data from Particle Data Group~\cite{ParticleDataGroup:2024cfk} 
for the central values of the following decay widths: 
\eq\label{Data_P2g}
\Gamma(\pi^0 \rightarrow \gamma\gamma) =  7.72 \, {\rm eV}
\,, \quad 
\Gamma(\eta \rightarrow \gamma\gamma)  =  0.516 \, {\rm keV}
\,, \quad 
\Gamma(\eta' \rightarrow \gamma\gamma) =  4.28 \, {\rm keV}
\,. 
\en
Using Eqs.~(\ref{decaygamma}) and~(\ref{Data_P2g}) we find
\eq\label{gP2g}
G_{\pi^0\gamma\gamma} = G_{\eta\gamma\gamma} = 0.274 \, {\rm GeV}^{-1}
\,, \qquad G_{\eta'\gamma\gamma} = 0.341 \, {\rm GeV}^{-1} \,. 
\en
Next, we can provide the limits on the couplings of
the partial contributions to the $G_{a\gamma\gamma}$ couplings.
One should stress that we did not assume that 
three possible contributions to the $G_{a\gamma\gamma}$ couplings 
have the same sign. As soon as the relative signs of three contributions
are not yet known, we can only derive the upper
limits on the magnitudes of the respective couplings. 

\section{Results}
\label{sec:results}

\subsection{Bounds on the PQ charges from invisible
  decays of pseudoscalar mesons $\pi^0$, $\eta$, and $\eta'$}

Using currently the best experimental upper limits 
for the branching ratios of the
$\pi^0, \eta, \eta' \rightarrow invisible$ decays:
\eq 
& &Br(\pi^0 \rightarrow invisible) \ < \ 4.4 \times 10^{-9} \,,
\label{piinvis}\\
& &Br(\eta \rightarrow invisible) \ < \ 1 \times 10^{-4} \,,
\label{etainvis}\\
& &Br(\eta' \rightarrow invisible) \ < \ 2.1 \times 10^{-4} \,,
\label{etaprimeinvis}
\en
respectively from the NA62~\cite{NA62:2020pwi},
BESIII~\cite{BESIII:2012nen}, and NA64~\cite{Andreev:2024tmh}
experiments,  
one can derive the bounds for the products of the couplings
involving $g_p$ and $g_s$ couplings of the ALP with fermion DM: 
\eq 
& &|\theta_{a\pi^0}| \ G_{\pi^0\chi} \ < \ 7.9 \times 10^{-8}
\label{boundpi}
\,,\\
& &|\theta_{a\eta}| \ G_{\eta\chi} \ < \ 1.9 \times 10^{-4}
\label{boundeta}
\,,\\
& &|\theta_{a\eta'}| \ G_{\eta'\chi} \ < \ 1 \times 10^{-3}
\,,
\label{boundetaprime} 
\en
where $G_{P\chi} = g_p \ \beta_{P\chi}^{1/2} $
or $G_{P\chi} = g_s \ \beta_{P\chi}^{3/2}$. 

For completeness we also consider the limiting case, when ALP reduces
to the QCD axion at $\Ma = 0$. In this case one present the upper limits
on the combinations of the PQ charges using
Eqs.~(\ref{boundpi})-(\ref{boundetaprime}) and 
Eq.~(\ref{theta_aP_exp}): 
\eq\label{PQcharges} 
& &|Q_d - Q_u| \ G_{\pi^0\chi} \ < \ 1.7 \times 10^{-3}
\ \frac{f_a}{{\rm 1 \, TeV}} \,,
\\
& &|Q_{\rm tot}| \ G_{\pi^0\chi} \ < \ 6.3 \times 10^{-3}
\ \frac{f_a}{{\rm 1 \, TeV}} \,,
\en
using upper limit for $|\theta_{a\pi^0}|$,
\eq
& &|Q_s| \ G_{\eta\chi} \ < \ 2.4 
\ \frac{f_a}{{\rm 1 \, TeV}} \,,
\\
& &|Q_{GQ}| \ G_{\eta\chi} \ < 71 
\ \frac{f_a}{{\rm 1 \, TeV}} \,,
\\
& &|Q_{\rm tot}| \ G_{\eta\chi} \ < 870 
\ \frac{f_a}{{\rm 1 \, TeV}} \,,
\en
using upper limit for $|\theta_{a\eta}|$,
and
\eq
& &|Q_{GQ}| \ G_{\eta'\chi} \ < \ 37 
\ \frac{f_a}{{\rm 1 \, TeV}} \,,
\\
& &|Q_{\rm tot}| \ G_{\eta'\chi} \ < \ 58  
\ \frac{f_a}{{\rm 1 \, TeV}} \,,
\en
using upper limit for $|\theta_{a\eta'}|$, where 
$Q_{\rm tot} = Q_u + Q_d + Q_s + Q_{GQ}$. 
Here and in the following we present the derived limits
on the mixing parameters $\theta_{aP}$ 
in terms of the coupling $f_a$ substituted in the units of TeV.
It is clear that increasing of $f_a$ leads to increasing of upper
limits for the products of combinations of the PQ charges and
couplings of ALP with DM fermions. 
Note similar estimates have been done before in Ref.~\cite{Ema:2020ulo}
in the framework of calculation without mixing of the pseudoscalar mesons.

Below, for completeness, we derive our estimates for the QCD axion mass using
upper limits for $|Q_{\rm tot}|$ derived from experimental
upper limits for invisible decay of $\pi^0$ 
\eq 
M_{a, {\rm QCD}} < \frac{2.7 \times 10^{-7}}{G_{\pi^0\chi}} \, M_{\pi} \,, 
\en 
of $\eta$
\eq 
M_{a, {\rm QCD}} < \frac{3.8 \times 10^{-3}}{G_{\eta\chi}} \, M_{\pi} \,, 
\en
of $\eta'$
\eq 
M_{a, {\rm QCD}} < \frac{2.5 \times 10^{-2}}{G_{\eta'\chi}} \, M_{\pi} \,.
\en 

\subsection{Search for ALPs in two photon decays}
  
The NA64 collaboration looked for ALPs using the reaction chain 
\begin{equation}
e^- Z  \rightarrow e^- Z \gamma; \gamma Z \to Z a; a \to \gamma\gamma
\label{eq:chain1}
\end{equation}
and obtained upper bound
$G_{a\gamma\gamma} < 2.2 \times 10^{-3}$ GeV$^{-1}$
for $M_a \leq 55$~MeV~\cite{NA64:2020qwq}. 

The number of events with pseudoscalars $P$ in 
the reaction \eqref{eq:chain1} is 
proportional to $G^2_{P\gamma\gamma}$. 
For the case of the invisible decays of $\pi^0$, $\eta$, and $\eta'$
the number of events with missing energy coming from
the invisible pseudoscalar decays is 
\eq
\label{pseudoinvisible} 
N_P(invisible ~decays) \sim G^2_{P\gamma\gamma}
\cdot {\rm Br}(P \rightarrow invisible) \,.
\en
By assumption ${\rm Br}(a \rightarrow invisible) = 1$ and taking
into account the mixing of ALP with pseudoscalar mesons
we can derive the upper limits for the branchings of invisible decays
of the pseudoscalar in the terms of the ratio of
two-photon couplings $G_{a\gamma\gamma}/G_{P\gamma\gamma}$
\eq 
{\rm Br}(P \rightarrow invisible)
< \biggl(\frac{G_{a\gamma\gamma}}{G_{P\gamma\gamma}}\biggr)^2 \,. 
\label{bound11}
\en
Using prediction of the NA64 experiment for
the upper bound of the ALP coupling $G_{a\gamma\gamma}$ and data for
the pseudoscalar couplings $G_{P\gamma\gamma}$
we find that
\eq 
\label{eta'inv} 
& &{\rm Br}(\pi^0 \rightarrow invisible) \ < \ 6.4 \times 10^{-5} \,,
\label{pi0inv}\\
& &{\rm Br}(\eta \rightarrow invisible) \ < \ 6.4 \times 10^{-5} \,,
\label{etainv}\\
& &{\rm Br}(\eta' \rightarrow invisible) \ < \ 4.2 \times 10^{-5} \,.
\en
So we see that obtained bounds for the ${\rm Br}(\eta \rightarrow invisible)$
and ${\rm Br}(\eta' \rightarrow invisible)$ are more stringent 
then data bounds~(\ref{etainvis}) and~(\ref{etaprimeinvis}).

As we discussed in the end of previous section, we can  
derived the upper limits on the magnitudes of the couplings
defining the partial contributions
to the $G_{a\gamma\gamma}$ couplings. In particular,
we can derive the upper limits for the mixing angles $|\theta_{aP}|$. 
In particular, using Eq.~(\ref{a2gfull}) and the upper bound for the coupling  
$G_{a\gamma\gamma} < 2.2 \times 10^{-3}$ GeV$^{-1}$~\cite{NA64:2020qwq}  
get the upper limits for $|\theta_{aP}|$, which are less stringent 
than ones derived from branchings of invisible decays  
\eq
|\theta_{a\pi^0}| <  8 \times 10^{-3} \,, \quad 
|\theta_{a\eta}| <  8 \times 10^{-3} \,, \quad 
|\theta_{a\eta'}| < 6.4 \times 10^{-3} \,. 
\en 
Also we derive the upper limits on the parameters of heavy quarks, which
define their contribution to the $G_{a\gamma\gamma}$ coupling:
\eq
\bigg|C_c \, \frac{M_a^2}{4 m_c^2}\bigg| =
\bigg|C_t \, \frac{M_a^2}{4 m_t^2}\bigg|
< 5 \,\,\frac{f_a}{{\rm 1 \, TeV}}\,,  \qquad
\bigg|C_b \, \frac{M_a^2}{4 m_b^2}\bigg| <
20 \,\,\frac{f_a}{{\rm 1 \, TeV}}\,.
\en 

\subsection{
  Search for ALPs in the chain reaction $\pi^- \,
  +  \, (Z,A) \, \rightarrow \,
  (Z-1, A)  \, +  \, a;  a \rightarrow invisible$}

In Ref.~\cite{Gninenko:2023rbf} we did detailed analysis
of the cross sections of charge exchange reaction
\eq 
\sigma(\pi^- \,+ (Z,A) \rightarrow P +\, (Z-1, A))  \,, \qquad
P = \pi^0, \eta, \eta' \,.
\en
In particular for Fe nuclei target we found~\cite{Gninenko:2023rbf}
that for final neutral light pseudoscalar mesons state
$P = \pi^0, ~\eta, ~\eta'$, the cross sections are equal
to 67.4, 21.9, 10.4~$\mu$b, respectively,
for the incoming pion energy 50 GeV.  

In this paper we propose a search for ALPs
in the chain reaction $\pi^- \,+ (Z,A) \, \rightarrow \, 
(Z-1, A) +a;~a \rightarrow invisible$. 
Using the mixing angles of ALP with pseudoscalars we derive
  the formula for the integral cross section
  of the ALP production in the charge-exchange reaction
  $\pi^- \,+ (Z,A) \rightarrow a \,+ (Z-1, A)$: 
\eq
\hspace*{-.3cm}
\sigma(\pi^- \,+ (Z,A) \rightarrow a \,+ (Z-1, A)) =
\sum_P \, \bar\theta^2_{aP} \ \sigma(\pi^- \, +(Z,A)
\rightarrow P \, +(Z-1, A)) \,.
\en
This formula allows to 
find bounds on the mixing parameters $ \bar\theta_{aP}$ and
then one can estimate the branchings of ALP invisible decays.
In particular for the ALP mass close to $\pi^0$ mass
the number of produced ALPs is given by the formula
\eq 
N_{a} = \pi\mathrm{OT}\times
\frac{\sigma(\pi^- \, +(Z,A) \rightarrow \pi^0 
\,+ (Z-1, A))}{\sigma_{\rm tot}} \, \theta_{a\pi}^2 \,,  
\en
where $\pi\mathrm{OT}$ is the number of pions on target. 
For the total $\pi^-$ scattering cross section on Fe nuclei equal to
$\sigma_{\rm tot}(\pi^- + {\rm Fe}) = A^{2/3} \sigma_{\rm tot}(\pi^- + p)
\sim 500$~mb at 50 GeV,  our estimate gives
\eq 
\frac{\sigma(\pi^-\, + (Z,A) \rightarrow \pi^0 + (Z-1, A))}
{\sigma_{\rm tot}} = {\cal O}(10^{-4}) \,.
\en
So for $ \pi\mathrm{OT} = 10^{12}$ in the assumption of the zero background 
we can expect to obtain the following estimation for the $|\theta_{a\pi^0}|$
\eq
|\theta_{a\pi^0}| = 10^{-2}  \,,
|\theta_{a\eta}|  = 10^{-2}  \,,
|\theta_{a\eta'}| = 10^{-3}  \,. 
\en 

\subsection{An estimate of the observable dark matter density}

We make standard assumption that in the early Universe dark matter
particles were in equilibrium with the SM particles.  
At some temperature decoupling of dark matter takes place 
that leads to the observable relic density of dark matter.
The solution of Boltzmann equation 
allows to estimate the cross section of 
the annihilation of dark fermions into quarks
$\chi \bar\chi \rightarrow q \bar q$.

We consider two possibilities of the annihilation
of dark fermions into quarks -- $S$-wave and $P$-wave,
which are described with the use of 
the pseudoscalar and scalar interaction Lagrangians of ALP
with dark fermions, respectively, derived in Eq.~(\ref{chia})  
\eq
\label{chia2}
{\cal L}_{a \chi \bar{\chi}} &=& \sum\limits_{\Gamma = s, p}
             {\cal L}_{a \chi \bar{\chi}}^{\Gamma}
\,,
\nonumber\\
{\cal L}_{a \chi \bar{\chi}}^{p} &=& 
g_{p} \, a \, \bar{\chi}  \, i \gamma_5  \, \chi
\,,
\quad 
{\cal L}_{a \chi \bar{\chi}}^{s} \ = \ 
g_{s} \, a \, \bar{\chi} \, \chi 
\en 
and pseudoscalar interaction Lagrangian of ALP with quarks
\eq
{\cal L}_{aq\bar q} = a 
\, \sum_{q = u, d, s} \, 
\bar q \, i \gamma_5 \, g_{aq} \, q
\,, 
\en
where $\chi$ is pseudo Dirac fermion and
$g_{aq}$ is the ALP-quark coupling. 
Here we consider the couplings of the ALP with
pseudoscalar currents composed of quarks and dark fermions,
respectively. 
The tree-level results for the annihilation cross section into
$u$, $d$, and $s$ quarks is for $S$- and $P$-wave transitions
are 
\eq 
\label{csformula1}
S-{\rm wave}: && \qquad
\sigma(\chi \bar{\chi} \rightarrow q \bar q) \, v_{rel}  =
\frac{3}{2\pi} 
\, g_{p}^2 \, \bar g_{aq}^2 
\, \frac{M^2_{\chi}}{(M^2_a - 4 M^2_{\chi})^2} \,,
\\
P-{\rm wave}: && \qquad
\sigma(\chi \bar{\chi} \rightarrow q \bar q) \, v_{rel}  =
\frac{3}{8\pi} 
\, g_{s}^2 \, \bar g_{aq}^2 
\, \frac{M^2_{\chi} \, v_{rel}^2}{(M^2_a - 4 M^2_{\chi})^2} \,.
\label{csformula2} 
\en
Here $v_{rel} = |\vec{v}_1 -  \vec{v}_2|$ is the relative velocity of
the annihilating DM particles and
$\bar g_{aq} = \sqrt{g^2_{au} +g^2_{au} + g^2_{as}}$. 

Numerically it is known~\cite{ParticleDataGroup:2024cfk} 
that the value of the thermal average of the annihilation cross section
times the relative velocity given by 
\eq 
\la\sigma(\chi \bar{\chi} \rightarrow \, SM \, particles) \, v_{rel}\ra 
= 2.6 \times 10^{-9}~{\rm GeV}^{-2}  
\label{csformulaSM}
\en
leads to observable DM density. 
Therefore,
\eq
& &\la\sigma(\chi \bar{\chi} \rightarrow q \bar q) \, v_{rel}\ra \,
= 2.6 \times 10^{-9}~{\rm GeV}^2 
\label{csformula.b} \,, \\
& &\la\sigma(\chi \bar{\chi} \rightarrow q \bar q) \, v_{rel}\ra \,
\frac{1}{\la v_{rel}^2\ra}
= 2 \times 10^{-8}~{\rm GeV}^2 \,, 
\label{csformula.a}
\en
where $\la v_{rel}^2\ra = 0.13$.

It should be noted that for the case
$2 M_{\chi} \leq 1$~GeV formulas~(\ref{csformula1})
and~(\ref{csformula2})
are  not quite correct since strong interaction effects are important. 
Using effective Lagrangians we can estimate
the hadronic effects with the  use of the $a$ mixing with pseudoscalars 
$\pi^0, ~\eta, ~\eta'$ and to estimate the correct value of
$\sigma(\chi \bar{\chi} \rightarrow hadrons)$.
Namely we assume that $\sigma(\chi \bar{\chi} \rightarrow hadrons) 
= \sigma(\chi \bar{\chi} \rightarrow a \rightarrow
(\pi^{0}, \eta, \eta') \rightarrow hadrons)$.  
Because the total widths of the $\pi^0$ and $\eta$ are much smaller
than the total width of the $\eta'$
\eq\label{totalwidths}
\Gamma_{\rm tot}(\eta') = 0.23 \, {\rm MeV} \gg
\Gamma_{\rm tot}(\eta)  = 1.31 \, {\rm keV} \gg
\Gamma_{\rm tot}(\pi^0) = 7.81 \, {\rm eV}  \,, 
\en   
the $\pi^0$ predominantly decays into two photons
with the fraction 98.8 \%, and the total fractions
of hadronic decay modes of the $\eta'$ (65 \%)
and $\eta$ (55.6 \%) are compatible we conclude that
the only $\eta'$ meson contributes to the cross section 
$\sigma(\chi \bar{\chi} \to \ a \to  hadrons)$. 
Due to this reason we shall consider the scenario that the 
double mass of the DM fermion lies in the interval 
$M_{\eta} < 2M_{\chi} < M_{\eta'}$ for which only invisible decays
$\eta' \rightarrow \chi \bar{\chi}$ are possible.
In Eq.~(\ref{totalwidths})~\cite{ParticleDataGroup:2024cfk}
we cite the central values of the averaged widths
of the $\eta'$, $\eta$, and $\pi^0$ from
Particle Data Group~\cite{ParticleDataGroup:2024cfk}. 

Formula for the determination of
annihilation cross section for the $S$- and $P$-wave
transitions reads 
\eq 
\la\sigma(\chi \bar{\chi} \rightarrow hadrons) \, v_{rel}\ra
&=& \frac{2 g^2_{p} M_{\chi}}{(M^2_a - 4 M^2_{\chi})^2}  \,  
\, \Gamma(\eta' \rightarrow hadrons) \, \bar\theta_{a\eta'}^2
\nonumber\\
&=& \frac{g^2_{s} M_{\chi} \, v_{rel}^2}{2 (M^2_a - 4 M^2_{\chi})^2}  \,  
\, \Gamma(\eta' \rightarrow hadrons) \,  
\bar\theta_{a\eta'}^2
\nonumber\\
&=& 2.6 \times 10^{-9}~{\rm GeV}^{-2} 
\,,
\label{csformula11}
\en
where $\Gamma(\eta' \rightarrow hadrons) \simeq
0.65 \, \Gamma_{\rm tot}(\eta') = 0.15$ MeV. 

As a numerical application we make the estimates for  
the ALP-quark coupling $\bar g_{aq}$ and the ALP-$\eta'$
mixing parameter $\bar\theta_{a\eta'}$ 
using Eqs.~(\ref{csformula1}), (\ref{csformulaSM}),
and~(\ref{csformula11}). 
The values $M_a$, $M_{\chi}$ are not known
and we consider three scenarios for the ratio  
of the ALP and DM fermion masses, 
$M_{a}/M_{\chi} = 2.5$,
$M_{a}/M_{\chi} = 5$, and
$M_{a}/M_{\chi} = 10$ with   
$M_{\chi} = 0.45 \, M_{\eta'}$.

For the $S$- and $P$-wave transitions 
we derive the following estimates
for the products
$|\bar g_{aq} \, g_p|$, 
$|\bar g_{aq} \, g_s|$,
$|\bar\theta_{a\eta'} \, g_p|$,
$|\bar\theta_{a\eta'} \, g_s|$: 
\begin{align} 
|\bar g_{aq} \, g_p| & = 7.2 \times 10^{-5} \,,
\hspace*{-2.5cm}
& \frac{M_{a}}{M_{\chi}} &= 2.5
\,,
\nonumber\\
|\bar g_{aq} \, g_p|
& = 6.7 \times 10^{-4} \,,
\hspace*{-2.5cm}
& \frac{M_{a}}{M_{\chi}} &= 5
\,,
\nonumber\\
|\bar g_{aq}| \, g_p|
& = 3.1 \times 10^{-3} \,,
\hspace*{-2.5cm}
& \frac{M_{a}}{M_{\chi}} &= 10 \,,
\end{align} 
\begin{align} 
|\bar g_{aq} \, g_s| &= 4 \times 10^{-4} \,,
\hspace*{-2.5cm}
& \frac{M_{a}}{M_{\chi}} &= 2.5
\,,
\nonumber\\
|\bar g_{aq} \, g_s|
& = 3.7 \times 10^{-3} \,,
\hspace*{-2.5cm}
& \frac{M_{a}}{M_{\chi}} &= 5
\,,
\nonumber\\
|\bar g_{aq} \, g_s|
& = 1.7 \times 10^{-2} \,,
\hspace*{-2.5cm}
& \frac{M_{a}}{M_{\chi}} &= 10 \,,
\end{align} 
\begin{align} 
|\bar\theta_{a\eta'} \, g_p| &= 2.8 \times 10^{-7} \,,
\hspace*{-2.5cm}
& \frac{M_{a}}{M_{\chi}} &= 2.5
\,,
\nonumber\\
|\bar\theta_{a\eta'} \, g_p| &= 2.6 \times 10^{-6} \,,
\hspace*{-2.5cm}
& \frac{M_{a}}{M_{\chi}} &= 5
\,,
\nonumber\\
|\bar\theta_{a\eta'} \, g_p| &= 1.2 \times 10^{-5} \,,
\hspace*{-2.5cm}
& \frac{M_{a}}{M_{\chi}} &= 10
\,,
\end{align} 
\begin{align} 
|\bar\theta_{a\eta'} \, g_s| &= 1.6 \times 10^{-6} \,,
\hspace*{-2.5cm}
& \frac{M_{a}}{M_{\chi}} &= 2.5
\,,
\nonumber\\
|\bar\theta_{a\eta'} \, g_s| &= 1.4 \times 10^{-5} \,,
\hspace*{-2.5cm}
& \frac{M_{a}}{M_{\chi}} &= 5
\,,
\nonumber\\
|\bar\theta_{a\eta'} \, g_s| &= 6.7 \times 10^{-5} \,,
\hspace*{-2.5cm}
& \frac{M_{a}}{M_{\chi}} &= 10
\,.
\end{align} 
From the invisible $\eta$ and $\eta'$ meson decays~(\ref{boundeta})
we get less stringent upper bounds on the product of couplings 
$|\bar\theta_{a\eta'} \, g_p|$ and $|\bar\theta_{a\eta'} \, g_s|$
than from annihilation
cross section of dark matter 
\begin{align} 
|\bar\theta_{a\eta'} \, g_p| &= 4.0 \times 10^{-4} \,,
\hspace*{-2.5cm} &\frac{M_{a}}{M_{\chi}} &= 2.5
\,,
\nonumber\\
|\bar\theta_{a\eta'} \, g_p| &= 5.2 \times 10^{-4} \,,
\hspace*{-2.5cm} &\frac{M_{a}}{M_{\chi}} &= 5
\,,
\nonumber\\
|\bar\theta_{a\eta'} \, g_p| &= 5.4 \times 10^{-4} \,,
\hspace*{-2.5cm} &\frac{M_{a}}{M_{\chi}} &= 10
\end{align} 
and
\begin{align} 
|\bar\theta_{a\eta'} \, g_p| &= 2.6 \times 10^{-4} \,,
\hspace*{-2.5cm} &\frac{M_{a}}{M_{\chi}} &= 2.5
\,,
\nonumber\\
|\bar\theta_{a\eta'} \, g_p| &= 4.6 \times 10^{-4} \,,
\hspace*{-2.5cm} &\frac{M_{a}}{M_{\chi}} &= 5
\,,
\nonumber\\
|\bar\theta_{a\eta'} \, g_p| &= 5.2 \times 10^{-4} \,,
\hspace*{-2.5cm} &\frac{M_{a}}{M_{\chi}} &= 10
\,.
\end{align} 

\subsection{Alternative method based on direct 
interaction of $\eta'$ with light dark matter particles}

Since we are mainly interested in the mass region
$M_{\eta} < 2 M_{\chi} < M_{\eta'}$
we can consider an effective interaction of $\eta'$ 
with $\chi \bar{\chi} $, namely
\eq 
   {\cal L}_{\eta' \chi\bar\chi} = \eta' \bar{\chi} \,
   (g_s + i \gamma_5 \, g_p) \, \chi \,,
\en
where $g_s = \theta_{\eta' a}g_{1\chi a}$,
      $g_p = \theta_{\eta' a}g_{2\chi a}$.
The decay width $\Gamma(\eta' \rightarrow \chi \bar{\chi})$
is given by the formula
\eq
\Gamma(\eta' \rightarrow \chi \bar{\chi}) =
\frac{M_{\eta'}}{8 \pi}(g_s^2 \, \beta^3_{\eta' \chi}
+  g_p^2 \, \beta_{\eta' \chi}) \,.
\en
As a consequence of experimental bound
$Br(\eta' \rightarrow invisible) < 2.1 \cdot10^{-4}$
and the value~(\ref{etaprimeinvis})  
$\Gamma_{tot}(\eta') = 0.23$~MeV we find that 
\eq 
|G_{\eta'\chi}| \ < \ 0.8 \times 10^{-3}
\,,
\label{Bound.G}
\en
where $G_{\eta'\chi} = g_p \ \beta_{\eta'\chi}^{1/2}$
or $G_{\eta'\chi} = g_s \ \beta_{\eta'\chi}^{3/2}$. 

The cross section for the $\chi \bar{\chi} \rightarrow hadrons$ based
on the use of the effective Lagrangian~(\ref{Lagr_Pchi}) reads
\eq 
\la\sigma(\chi \bar{\chi} \rightarrow hadrons) \, v_{rel}\ra =
      \frac{g^2_p (1 - \beta_{\eta'\chi}^2)}
      {M_{\eta'}^3 \, \beta_{\eta'\chi}^4} \, 
      \Gamma_{\rm tot}(\eta') 
\en
for $g_s = 0$ and
\eq 
\la\sigma(\chi \bar{\chi} \rightarrow hadrons) \, v_{rel}\ra 
= \frac{g^2_s (1 - \beta_{\eta'\chi}^2) \, v^2_{rel}}
      {4 \, M_{\eta'}^3 \, \beta_{\eta'\chi}^4} \, 
      \Gamma_{\rm tot}(\eta')
\en
for $g_p = 0$.
In our estimates we take the values~(\ref{csformulaSM}) and~(\ref{csformula.a})
and we find 
\eq 
\frac{g^2_P(1 - \beta^2_{\eta' \chi})}{\beta^4_{\eta' \chi}}
= 1.2 \times 10^{-5}\,, \qquad
\frac{g^2_S(1 - \beta^2_{\eta' \chi})}{\beta^4_{\eta' \chi}} = 3.7 \times 10^{-4} \,. 
\en   
As a consequence of the inequality~(\ref{Bound.G}) we obtain 
\eq 
\beta_{\eta' \chi} < 0.629\,, \quad M_{\chi} > 0.39 M_{\eta'} 
\label{Bound.a}
\en
for pseudoscalar coupling  ($g_p \neq 0,~g_s = 0$) and 
\eq 
\beta_{\eta' \chi} < 0.454 \,, \quad  M_{\chi} > 0.44 M_{\eta'}
\label{Bound.b}
\en
for scalar coupling $(g_s \neq 0,   ~g_p =  0)$. 
Note that we assumed that $M_{\chi} < 0.5 M_{\eta'}$
for scalar coupling $(g_s \neq 0,   ~g_p =  0)$. 
So we see that the obtained bounds on $M_{\chi}$ are very strong and lead to 
some fine tuning in the choose of $M_{\chi}$.
Due to resonance type of the annihilation cross section
the improvement for the ${\rm Br}(\eta' \rightarrow invisible)$ bound
will make the bounds~(\ref{Bound.a}) and~(\ref{Bound.b}) more stringent.

\section{Summary}
\label{sec:summary}

In this paper, we discuss leptophobic dark sector with
the pseudoscalar portal involving ALPs particles.
In particular, we include the mixing of the ALP with light
pseudoscalar mesons $P = \pi^0, \eta, \eta'$ and derive the new
limits on the couplings of the ALP with quarks
and the couplings of  $P$ states with dark fermions using
upper limits for the branchings of the invisible modes
${\rm Br}(P \rightarrow invisible)$ and constraints 
obtained by the NA64 experiment at CERN. In our study we
establish the limits on the ALP-pseudoscalar mesons
mixing parameters in the presence
of the term explicitly breaking the PQ symmetry. 
In our numerical analysis we derived upper limits on the mixing
parameters of ALP with pseudoscalars $P$ using current data on
invisible decays and two-photon decays of the pseudoscalar
mesons.

Novel idea proposed in our paper is conjecture that
due to mixing with pseudoscalar mesons massive
ALP can decay via invisible modes into fermionic Dark Matter.
We derive corresponding Lagrangian and derive the expression for
the cross section of ALP production in the charge-exchange
reactions of negative pion on nuclei targets.
Note, invisible decays of massive ALP 
are not kinematically forbidden in comparison with very light
QCD axion. NA64h setup of the  
NA64 Collaboration~\cite{Andreev:2024tmh} gives the 
opportunity for study of invisible decays of ALPs.
From study of these reactions one can derive
upper limits on mixing parameters of ALP with pseudoscalar mesons. 
We did detailed numerical analysis of ALP parameters
(mixing parameters and coupling constants with dark fermions)
using invisible decays of light pseudoscalars $(\pi^0, \eta, \eta')$,
production of ALP in charge-exchange reactions,
two-photon decays, annihilation of dark fermions into quarks.
Also we made the predictions for the
partial contributions of the PQ charges
the ALP-pseudoscalar mesons mixing 
parameters as functions of the ALP mass~$M_a$.

\begin{acknowledgments} 

This work was funded
by FONDECYT (Chile) under Grant No. 1240066, and
by ANID$-$Millen\-nium Program$-$ICN2019\_044 (Chile). The work of A.~S.~Zh. 
is supported by Russian Science Foundation (grant No. RSF 23-22-00041). 

\end{acknowledgments}

\appendix
\section{ALP-pseudoscalar mesons mixing and diagonalization of mass term}
\label{App:A}

In this Appendix we present details of diagonalization of the mass term
involving the bare ALP field $a^0$ and neutral 
pseudoscalars $P=(\pi^3, \eta^8, \eta^0)$ leading to
mixing of physical states $a$, $\pi^0$, $\eta$, $\eta'$.

First, we discuss details of diagonalization for the case
when $\Ma \neq 0$ and consider limit $\Ma \to 0$, i.e.
when ALP reduces to the QCD axion. 
As we stressed before the ALP as much heavier
state than QCD axion and there is a strong interest for searching 
for invisible decays of ALP at the NA64h setup of the
NA64 experiment. 

The part of the underlying Lagrangian~(\ref{ALP_P_Lagrangian})
relevant for the mixing of ALP and pseudoscalar mesons  
(see also Ref.~\cite{Alves:2024dpa})  
\eq\label{ALP_P_mixing}
{\cal L}_{\rm ALP+P}^{\rm mix} = \frac{1}{2} \Phi_0^T \,
{\cal M}^2_{\Phi_0} \, \Phi_0 \,,
\en
where $\Phi_0^T = (a^0, \pi^3, \eta^8, \eta^0)$ is the quadruplet
of the bare mixing states and 
$M^2_{\Phi_0}$ is the $4 \times 4$ mass mixing matrix 
\eq
 {\cal M}^2_{\Phi_0} = 
\begin{pmatrix}
  M_{\pi^3}^2 & 
  \  \ M_{\pi^3\eta^8}^2 &
  \  \ M_{\pi^3\eta^0}^2 &
  \  \ M_{a^0\pi^3}^{2} \\[2mm]
  M_{\pi^3\eta^8}^2 &
  \  \ M_{\eta^8}^2 &  
  \  \ M_{\eta^8\eta^0}^2 &
  \  \ M_{a^0\eta^8}^{2} \\[2mm]
  M_{\pi^3\eta^0}^2 &
  \  \ M_{\eta^8\eta^0}^2 &
  M_{\eta^0}^2 &  
  \  \ M_{a^0\eta^0}^{2} \\[2mm]
  M_{a^0\pi^3}^2 &
  \  \ M_{a^0\eta^8}^2 &
  \  \ M_{a^0\eta^0}^2 &  
  \  \ M_{a^0}^{2} \\
\end{pmatrix}\,,
\label{M2_mixing}
\en
where for convenience we denote the bare ALP as $a^0$. 
The elements of the mixing matrix
read~\cite{Weinberg:1978kz,Gasser:1983yg,Gasser:1984gg,Leutwyler:1997yr,Leutwyler:1996np,Alves:2024dpa}
\eq 
M_{\pi^3}^2  &=& B \, (m_u + m_d) \,, \nonumber\\
M_{\eta^8}^2 &=& \frac{B}{3} \, (m_u + m_d + 4m_s) \,, \nonumber\\
M_{\eta^0}^2 &=& M_0^2 + \frac{2 B}{3} \, (m_u + m_d + m_s) \,, \nonumber\\
M_{a^0}^2   &=& \Ma^2 + \frac{F^2}{f_a^2}  \, B \,
\Big(m_u Q_u^2 + m_d Q_d^2 + m_s Q_s^2 + \frac{M_0^2}{6 B} \, Q_{GQ}^2 
\Big) \,, \nonumber\\
M_{\pi^3\eta^8}^2 &=& \frac{M_{\pi^3\eta^0}^2}{\sqrt{2}} \ = \
\frac{B}{\sqrt{3}} \, (m_u - m_d)
\,, \nonumber\\
M_{\eta^8\eta^0}^2 &=& \frac{\sqrt{2}}{3} \, B (m_u + m_d - 2 m_s)
\,, \nonumber
\en
\eq
M_{a^0\pi^3}^2 &=& \frac{F}{f_a}  \, B  \, (m_u Q_u - m_d Q_d)
\,, \nonumber\\
M_{a\eta^8}^2  &=& \frac{F}{f_{a}}  \, \frac{B}{\sqrt{3}}
\, (m_u Q_ u + m_d Q_d - 2 m_sQ_s)
\,, \nonumber\\
M_{a\eta^0}^2  &=& \frac{F}{f_{a}}  \,  \sqrt{\frac{2}{3}} \, B \,
\Big(m_u Q_u + m_d Q_d + m_s Q_s - \frac{M_0^2}{2 B} \, Q_{GQ} 
\Big) \,.
\en
Diagonalization of the mixing term~(\ref{ALP_P_mixing}) is performed
with the use of the $4 \times 4$ orthogonal
matrix $\mathds{T}(\Phi)$~\cite{Choi:1986zw,Aloni:2018vki,Alves:2024dpa},
which is conventionally can be factorized as the product of two other
$4 \times 4$ matrices $\mathds{T}(a)$ and $ \mathds{T}(P)$ as
$\mathds{T}(\Phi) = \mathds{T}(a) \, \mathds{T}(P)$.
Here, first the matrix $\mathds{T}(a)$ removes the mixing
of the ALP with pseudoscalars and parametrized in terms of
the corresponding mixing
angles $\theta_{aP}$ with $P = \pi^3, \eta^8, \eta^0$ and
then the matrix $\mathds{T}(P)$
performs diagonalization in the sector of the neutral pseudoscalars.
Based on this,
the transformation between the quadruplets of
bare $\Phi_0 = (\pi^3, \eta^8, \eta^0, a^0)$ 
and physical $\Phi^T = (\pi^0, \eta, \eta', a)$ states reads:
\eq 
\Phi_0 = \mathds{T}(a)  \, \mathds{T}(P)  \, \Phi  
\,, 
\en 
where 
\eq
 \mathds{T}(a) = 
\left(
\begin{array}{ccccc}
&   &   &  \theta_{a\pi^0}                         \\[2mm]
&   \mathds{1}_{3\times3} &  & \theta_{a\eta}       \\[2mm]
&   &  & \theta_{a\eta'}                           \\[2mm]
- \theta_{a\pi^0} & - \theta_{a\eta} & - \theta_{a\eta'} & 1 
\end{array}
\right)\,, \qquad 
\mathds{T}(P) =
\left(
\begin{array}{ccccc}
&  &   &                    \ \ 0   \\[2mm]
&  \mathds{P}_{3\times3} &  & \ \ 0   \\[2mm]
&   &  &                     \ \ 0   \\[2mm]
0   &  0  &  0 &              \ \ 1 
\end{array}
\right)
\,,
\label{Eq:Mixing}
\en
where $\mathds{P}_{3\times3}$ is the $3 \times 3$ orthogonal matrix diagonalizing
remaining terms involving the triplet of neutral pseudoscalars to first order
in isospin breaking~\cite{Leutwyler:1996np,Alves:2024dpa,Escribano:2016ntp}:
\eq 
\mathds{P} =
\begin{pmatrix}
1 & \  \ - \theta_{\pi^0\eta} & \  \ - \theta_{\pi^0\eta'}\\[2mm]
   \theta_{\pi^0\eta}\cos\theta_{\eta\eta'}
   + \theta_{\pi^0\eta'}\sin\theta_{\eta\eta'} & 
   \  \ \cos\theta_{\eta\eta'} & \  \ \sin\theta_{\eta\eta'}\\[2mm]
   \  \ \theta_{\pi^0\eta'}\cos\theta_{\eta\eta'}
   -\theta_{\pi^0\eta}\sin\theta_{\eta\eta'}    &
   \  \ - \sin\theta_{\eta\eta'} & \  \ \cos\theta_{\eta\eta'}\\
\end{pmatrix}
\,.
\en
Here, 
$\theta_{\pi^0\eta}$, $\theta_{\pi^0\eta'}$, $\theta_{\eta\eta'}$ 
are the corresponding mixing angles. 

Finally, the sum of the mass terms of the ALP and neutral
light pseudoscalar mesons
($\pi^0$, $\eta$ and $\eta'$) reads 
\eq\label{ALP_P_diag}
{\cal L}_{\rm ALP+P}^{\rm mass} =
\frac{1}{2} \Phi^T \, {\cal M}^2_{\Phi} \, \Phi \,, \qquad 
{\cal M}^2_{\Phi} = {\rm diag}(M_{\pi^0}^2,M_{\eta}^2,M_{\eta'}^2,M_{a}^2)
\,. 
\en
We list the explicit results for the mixing
angles $\theta_{aP_i}$, $\theta_{P_1P_2}$ and
masses $M_{\pi^0}^2$, $M_{\eta}^2$, $M_{\eta'}^2$,
and $M_{a}^2$ below. 

To first order in isospin breaking and $F/f_a$ expansion,
the relations between the bare states ($\pi^3$, $\eta^8$, $\eta^0$, $a^0$)
and the mass eigenstates ($\pi^0$, $\eta$, $\eta'$, $a$) are 
\eq
\hspace*{-.5cm}
\pi^3 &=&
  \pi^0
- \theta_{\pi^0\eta}  \, \eta
- \theta_{\pi^0\eta'} \, \eta'
+ \theta_{a\pi^0}   \, a \,, 
\nonumber\\
\hspace*{-.5cm}
\eta^8 &=&
  \cos\theta_{\eta\eta'} \, \eta
+ \sin\theta_{\eta\eta'} \, \eta' 
+ \Big(\theta_{\pi^0\eta}  \, \cos\theta_{\eta\eta'}
     + \theta_{\pi^0\eta'} \, \sin\theta_{\eta\eta'}\Big) \, \pi^0
+ \theta_{a\eta} \, a \,, 
\nonumber\\
\hspace*{-.5cm}
\eta^0 &=&
  \cos\theta_{\eta\eta'} \, \eta'
- \sin\theta_{\eta\eta'} \, \eta 
+ \Big(\theta_{\pi^0\eta'} \, \cos\theta_{\eta\eta'}
     - \theta_{\pi^0\eta}  \, \sin\theta_{\eta\eta'}\Big) \, \pi^0
+ \theta_{a\eta'} \, a \,, 
\nonumber\\
\hspace*{-.5cm}
a^0 &=&
  a
  - \theta_{a\pi^0} \, \pi^0
  - \Big(\theta_{a\eta}  \, \cos\theta_{\eta\eta'}
     - \theta_{a\eta'} \, \sin\theta_{\eta\eta'}\Big) \, \eta
- \Big(\theta_{a\eta'} \, \cos\theta_{\eta\eta'}
     + \theta_{a\eta}  \, \sin\theta_{\eta\eta'}\Big) \, \eta'  \,.
\en
First, we specify the mixing angles and the masses for the
subsystem of the pseudoscalar mesons
($\pi^0$, $\eta$, $\eta'$)~\cite{Leutwyler:1996np,Leutwyler:1997yr}. 
As the result of the diagonalization of the mass mixing term
for the mixing angle $\theta_{\eta\eta'}$ two independent
Leutwyler relations have been derived in Ref.~\cite{Leutwyler:1996np}: 
\eq\label{Leutwyler_eqs} 
\sin 2\theta_{\eta\eta'} &=& \frac{2 M_{\eta^8\eta^0}^2}{M_{\eta'}^2-M_{\eta}^2}
= - \frac{4 \sqrt{2}}{3} \, \frac{M_K^2-M_{\pi}^2}{M_{\eta'}^2-M_{\eta}^2} \,,
\nonumber\\
\tan\theta_{\eta\eta'}   &=& \frac{M_{\eta^8}^2-M_{\eta}^2}{M_{\eta^8\eta^0}}
= - \frac{3}{2 \sqrt{2}} \, \frac{M_{\eta^8}^2-M_{\eta}^2}{M_K^2-M_{\pi}^2} \,,
\en
where $M_K^2 = (M_{K^+}^2 + M_{K^0}^2)/2 = B (m_s + \hat{m})$,
$M_{\pi}^2 =  (M_{\pi^+}^2 + M_{\pi^0}^2)/2 = 2 B \hat{m}$,
and $\hat{m} = (m_u + m_d)/2$.  

Using Eq.~(\ref{Leutwyler_eqs}) one can derive two additional
useful relations~\cite{Leutwyler:1997yr}:
\eq\label{Leutwyler_eqs2}
\sin^2\theta_{\eta\eta'} &=&
\frac{M_{\eta^8}^2-M_{\eta}^2}{M_{\eta'}^2-M_{\eta}^2}
\,, \nonumber\\
\Big(M_K^2-M_{\pi}^2\Big)^2 &=& \frac{9}{8} \,
\Big(M_{\eta'}^2-M_{\eta^8}^2\Big) \Big(M_{\eta^8}^2-M_{\eta}^2\Big) 
\,.
\en
For convenience we perform expansion 
of the masses and mixing parameters
in powers of the parameter
$x = B m_s/M_0^2 = (2 M_K^2 - M_{\pi}^2)/(2 M_0^2)$,
which is the expansion parameter in the large $N_c$ ChPT
encoding explicit breaking of $SU_L(3)  \times SU_R(3)$
chiral symmetry~\cite{Leutwyler:1997yr}.
Taking into account
that $x^2 \sim 0.1$ below we display results for the
expansion of the mentioned quantities up second order in $x$,
while we can do it up to any desired order in $x$.

In particular, the expansion for
the $\sin 2\theta_{\eta\eta'}$ starts from $x$.
Keeping the first oder in $\hat{m}/m_s$
and two orders in $x$ we get: 
\eq\label{sin2e_exp}
\sin 2\theta_{\eta\eta'} &=&
- \frac{4 \sqrt{2}}{3} \, x \,  
\biggl[
1 + \frac{2}{3} x + \frac{4}{9} x^2
- \frac{\hat{m}}{m_s} \biggl(
1 + \frac{4}{3} x + \frac{4}{3} x^2
\biggr) + {\cal O}\biggl(x^3,\frac{\hat{m}^2}{m_s^2}\biggr) 
\biggr]
\,.
\en
Using first Equation from~(\ref{Leutwyler_eqs}) 
and experimental values for the pion and kaon masses
$M_{\pi^+} = 0.13957$ GeV, $M_{\pi^0} = 0.1349768$ GeV,
$M_{K^+} = 0.493677$ GeV, $M_{K^0} = 0.497611$ GeV 
we fix $\theta_{\eta\eta'} \simeq - 21.9^\circ$. This prediction 
is in very good agreement with the improved measurements made
recently by the BESIII
Collaboration~\cite{BESIII:2023fai}:
$\theta_{\eta\eta'} \simeq - (22.11 \pm 0.26)^\circ$. 
To guarantee that Eq.~(\ref{sin2e_exp}) gives the same prediction
for the $\theta_{\eta\eta'} \simeq - 21.9^\circ$ as 
first equation in~(\ref{Leutwyler_eqs}) we can fix our free parameter
$M_0$ as  $M_0 = 0.857$ GeV. In Eq.~(\ref{sin2e_exp}) we use
canonical value of ChPT for the
ratio of quark masses $m_s/\hat{m} = 25$~\cite{Weinberg:1978kz,Gasser:1983yg,Gasser:1984gg}.
It gives the value of the expansion parameter in $x = 0.309$. 

For completeness we also present the expansion for 
the $\cos\theta_{\eta\eta'}$ and $\sin\theta_{\eta\eta'}$
\eq\label{cose_sine_exp}
\cos\theta_{\eta\eta'} &=&
1 - \frac{4}{9} x^2
+ \frac{8}{9} \frac{\hat{m}}{m_s} \, x^2
 + {\cal O}\biggl(x^3,\frac{\hat{m}^2}{m_s^2}\biggr) 
\biggr]
\,, \nonumber\\
\sin\theta_{\eta\eta'} &=&
- \frac{2 \sqrt{2}}{3} \, x \,  
\biggl[
1 + \frac{2}{3} x + \frac{8}{9} x^2
- \frac{\hat{m}}{m_s} \biggl(
1 + \frac{4}{3} x + \frac{8}{3} x^2
\biggr) + {\cal O}\biggl(x^3,\frac{\hat{m}^2}{m_s^2}\biggr) 
\biggr]
\,.
\en
Next we specify the mixing angles of pion with $\eta$
and $\eta'$ 
\eq
\theta_{\pi^0\eta}  &=& A_{\pi^0\eta\eta'} \, \cos\theta_{\eta\eta'}
                   -  B_{\pi^0\eta\eta'} \, \sin\theta_{\eta\eta'}
\,,
\nonumber\\
\theta_{\pi^0\eta'} &=& B_{\pi^0\eta\eta'} \, \cos\theta_{\eta\eta'}
                   +  A_{\pi^0\eta\eta'} \, \sin\theta_{\eta\eta'}
\,,
\en
where
\eq
A_{\pi^0\eta\eta'} &=&
\frac{M_{\pi^3\eta^8}^2 (M_{\eta^0}^2 - M_{\pi^3}^2)
   -  M_{\eta^8\eta^0}^2  M_{\pi^3\eta^0}^2}
     {M_{\eta^8\eta^0}^4 - (M_{\eta^0}^2
    - M_{\pi^3}^2) (M_{\eta^8}^2 - M_{\pi^3}^2)} \,,
\nonumber\\
B_{\pi^0\eta\eta'} &=&
\frac{M_{\pi^3\eta^0}^2 (M_{\eta^8}^2 - M_{\pi^3}^2)
   -  M_{\eta^8\eta^0}^2  M_{\pi^3\eta^8}^2}
     {M_{\eta^8\eta^0}^4 - (M_{\eta^0}^2
    - M_{\pi^3}^2) (M_{\eta^8}^2 - M_{\pi^3}^2)} \,.
\en
Expansion for the $\theta_{\pi^0\eta}$, $\theta_{\pi^0\eta'}$,
and their ratio $R_{\eta\eta'} = \theta_{\pi^0\eta'}/\theta_{\pi^0\eta}$ reads 
\eq
\theta_{\pi^0\eta} &=& \frac{\sqrt{3}}{4} \, \frac{\delta m}{m_s} \, 
\biggl[
1 + 2 x + \frac{20}{9} x^2 
+ \frac{\hat{m}}{m_s} \, \biggl(
1 - \frac{20}{9} x^2 
\biggr) + {\cal O}\biggl(x^3,\frac{\hat{m}^2}{m_s^2}\biggr) 
\biggr]
\,, \nonumber\\
\theta_{\pi^0\eta'} &=& \sqrt{\frac{2}{3}} \, x \, \frac{\delta m}{m_s} \, 
\biggl[
1 - \frac{4}{3} x - \frac{16}{9} x^2
+ \frac{\hat{m}}{m_s} \, \biggl(
\frac{4}{3} x + \frac{32}{3} x^2 
\biggr) + {\cal O}\biggl(x^3,\frac{\hat{m}^2}{m_s^2}\biggr) 
\biggr] 
\,,
\en
where $\delta m = m_d - m_u$ is the $d - u$ quark mass difference,
which encodes the strong isospin breaking effect. 
Using canonical values of the $\delta m = 4$ MeV we get:
\eq
\theta_{\pi^0\eta} = 0.018 \,, \qquad \theta_{\pi^0\eta'} = 0.0026 \,.
\en
One can see that the $\theta_{\pi^0\eta'}$ mixing parameter
is suppressed by one order in comparison with the parameter
$\theta_{\pi^0\eta}$.

Expressions for the masses of the pseudoscalar mesons after diagonalization
and their expansions are given by 
\eq 
M_{\pi^0}^2 &=& M_{\pi^3}^2
 + 2 M_{\pi^3\eta^8}^2 \, \Big(\theta_{\pi^0\eta}  \cos\theta_{\eta\eta'} 
 + \theta_{\pi^0\eta'} \sin\theta_{\eta\eta'}\Big)
 + 2 M_{\pi^3\eta^0}^2 \, \Big(\theta_{\pi^0\eta'} \cos\theta_{\eta\eta'}
 - \theta_{\pi^0\eta} \sin\theta_{\eta\eta'}\Big)  
\nonumber\\[1mm]
&=& 2 \, B \, \hat{m} \, \biggl[1 -  \frac{\delta m^2}{4 \, \hat{m} \, m_s} \, 
   \Big(1 + \frac{2}{3} x + \frac{32}{3} x^2\Big)
   + {\cal O}\Big(x^3,\frac{\delta m \, \hat{m} }{m_s^2},
   \frac{\delta m^2}{m_s^2}\Big)\biggr] 
\,, \nonumber\\[1mm]
M_{\eta}^2 &=& M_{\eta^8}^2 \, \cos^2\theta_{\eta\eta'} 
           +  M_{\eta^0}^2 \, \sin^2\theta_{\eta\eta'}
           -  M_{\eta^8\eta^0}^2 \, \sin 2\theta_{\eta\eta'}
\nonumber\\[1mm]
&=& \frac{4}{3} \, B \, m_s \, \biggl[1 - \frac{2}{3} x - \frac{4}{9} x^2
  +  \frac{\hat{m}}{m_s} \, \Big(\frac{1}{2}
  + \frac{4}{3} x + \frac{4}{3} x^2\Big)
           + {\cal O}\Big(x^3,\frac{\hat{m}^2}{m_s^2}\Big)\biggr] 
\,, \nonumber\\[1mm]
M_{\eta'}^2&=& M_{\eta^0}^2 \, \cos^2\theta_{\eta\eta'} 
           +  M_{\eta^8}^2 \, \sin^2\theta_{\eta\eta'}
           +  M_{\eta^8\eta^0}^2 \, \sin 2\theta_{\eta\eta'}
\nonumber\\[1mm]
&=&  \frac{2}{3} \, B \, m_s \, \biggl[\frac{3}{2 x}
 + 1 + \frac{4}{3} x + \frac{8}{9} x^2
 +  \frac{\hat{m}}{m_s} \, \Big(2 - \frac{8}{3} x - \frac{8}{3} x^2\Big)
           + {\cal O}\Big(x^3,\frac{\hat{m}^2}{m_s^2}\Big)\biggr] 
\,. 
\en 
Below we list results for the mass of ALP squared $M_A^2$
and ALP-pseudoscalar mesons mixing parameters $\theta_{aP}$: 
\eq\label{theta_aP_full}
\theta_{a\pi^0}^{\rm full} &=& \frac{F}{f_a} \, B \hat{m} \, \frac{I_{a\pi}}{J}
\,, \nonumber\\[1mm]
\theta_{a\eta}^{\rm full}
&=& \frac{F}{f_a \sqrt{3}} \, B \hat{m} \, \frac{I_{a\eta}}{J}
\,, \nonumber\\[1mm]
\theta_{a\eta'}^{\rm full}
&=& \frac{F}{f_a \sqrt{6}} \, B \hat{m} \, \frac{I_{a\eta'}}{J}
\,,
\en
where $I_{aP}$ and $J$ are given by  
\eq
\label{IaP}
I_{aP} &=&   \alpha_1^{P} \, \Ma^4
         +  \alpha_2^{P} \, \Ma^2 \, B m_s
         +  \alpha_3^{P} \, (B m_s)^2
\,, \nonumber\\[1mm]         
J    &=&    \beta_1 \, \Ma^6
        +   \beta_2 \, \Ma^4 \, B m_s
        +   \beta_3 \, \Ma^2 \, (B m_s)^2 
        +   \beta_4 \, (B m_s)^3 
\en
and 
\eq
\label{theta_aP_full2}
\alpha_1^{\pi^0} &=&
x  \biggl[Q_d - Q_u +  \frac{\delta m}{2 \hat{m}}  \, (Q_u + Q_d)\biggr]
\,, \nonumber\\[1mm]
\alpha_2^{\pi^0} &=&
- (Q_d - Q_u) \biggl[1 + 2 x +  \frac{4 x \, \epsilon}{1 + 6 x} 
\biggr] - \frac{\delta m}{\hat{m}} \, (Q_u + Q_d) \, (2+ x)
- \frac{\delta m}{3 \hat{m}} \, Q_{GQ} 
\,, \nonumber\\[1mm]
\alpha_3^{\pi^0} &=& \frac{4}{3}  \biggl[(Q_d - Q_u) (1 + \epsilon)
  + \frac{\delta m}{2 \hat{m}} \, Q_{\rm tot}\biggr]
\,, \nonumber\\[1mm]
\alpha_1^{\eta} &=& \frac{2 m_s}{\hat{m}} x 
\biggl[Q_s - \frac{\hat{m}}{2 m_s} (Q_u + Q_d) - \frac{\delta m}{4 m_s}
\, (Q_d - Q_u)\biggr]
\,, \nonumber\\[1mm]
\alpha_2^{\eta} &=&
\frac{2 m_s}{\hat{m}} \biggl[- Q_s \Big(1 + 4 x \frac{\hat{m}}{m_s}\Big) 
  - \frac{Q_{GQ}}{3} \Big(1 - \frac{\hat{m}}{m_s}\Big)
  + \frac{\hat{m}}{2m_s} (Q_u + Q_d) \Big(1 + 2 x
  + \frac{4 x  \, \epsilon}{1 + 6 x}\Big)
\nonumber\\[1mm]
&+& \frac{\delta m}{4 m_s} (Q_d - Q_u) (1 + 2 x) 
\biggr] 
\,, \nonumber\\[1mm]
\alpha_3^{\eta} &=& 4 \biggl[\Big(Q_s + \frac{Q_{GQ}}{3}\Big) (1 + \epsilon) 
  - Q_{\rm tot}  \frac{1 + 2 x}{1 + 6 x} \, \epsilon\Big)
\biggr]
\,, \nonumber\\
\alpha_1^{\eta'} &=& \frac{m_s}{\hat{m}} 
\biggl[Q_{GQ} - 2 x \Big(Q_s + \frac{\hat{m}}{m_s}
(Q_u + Q_d)
+ \frac{\delta m}{2 m_s} (Q_d - Q_u)\Big)
\biggr]
\,, \nonumber\\[1mm]
\alpha_2^{\eta'} &=& \frac{m_s}{\hat{m}}
\biggl[- \frac{4}{3} Q_{GQ} \Big(1 + \frac{2 \hat{m}}{m_s}\Big) 
  + 4 x \frac{\hat{m}}{m_s} \Big(2 Q_s + (Q_u + Q_d)
  \Big(1 + \frac{2 \epsilon}{1 + 6 x}\Big)\Big)
  + 2 x \frac{\delta m}{m_s} (Q_d - Q_u) 
\biggr] 
\,, \nonumber\\[1mm]
\alpha_3^{\eta'} &=&  \frac{8}{3} \biggl[Q_{GQ} (1 + \epsilon)
  - Q_{\rm tot}  \frac{6 x \, \epsilon}{1 + 6 x} 
  \biggr]
\,, \nonumber\\[1mm]
\beta_1 &=& -x
\,, \nonumber\\[1mm]
\beta_2 &=& 1 + 2 x \Big(1 +  2 \frac{\hat{m}}{m_s}\Big)
\,, \nonumber\\[1mm]
\beta_3 &=& -
\frac{4}{3} \Big(1 + 2 \frac{\hat{m}}{m_s}
(1 + 3 x + \frac{3 x \, \epsilon}{1 + 6 x}\Big)
\,, \nonumber\\[1mm]
\beta_4 &=& \frac{8}{3} \frac{\hat{m}}{m_s} (1 + \epsilon) 
\,. 
\en
We should note that the ALP mass in the full scheme is defined as
\eq\label{MA2_full}
M_a^2 
&=& \Ma^2 +
\theta_{a\pi^0}  \, \Big(\theta_{a\pi^0} M_{\pi^0}^2 + 2 M_{a^0\pi^3}\Big) 
\nonumber\\[1mm]
&+&
\theta_{a\eta}  \, \Big(\theta_{a\eta} M_{\eta}^2 + 2 M_{a^0\eta^8}\Big) 
+ \theta_{a\eta'}  \, \Big(\theta_{a\eta'} M_{\eta'}^2 + 2 M_{a^0\eta^0}\Big) 
\nonumber\\[1mm]
&+&2 \, \theta_{a\eta}  \, \theta_{a\eta'}      \, M_{\eta^8\eta^0}^2
 + 2 \, \theta_{a\pi^0} \, \Big(\theta_{a\eta}  \, M_{\pi^3\eta^8}
                             + \theta_{a\eta'} \, M_{\pi^3\eta^0}\Big)  
\,.
\en
For completeness, we present results for the same parameters
in the limit $\Ma = 0$, i.e. when ALP reduces to the QCD axion
(see also Ref.~\cite{Alves:2024dpa}).
In comparison with Ref.~\cite{Alves:2024dpa}
we additionally present the chiral and isospin-breaking expansion
of the QCD axion mass and its mixing parameters with pseudoscalar mesons.
We obtain: 
\eq
\label{theta_aP_exp}
   \stackrel{0}{\theta}_{a\pi^0}
   &=& \frac{F}{2 f_a} 
  \, \biggl[Q_d - Q_u 
    + \frac{\delta m}{2 \hat{m}} \, \frac{Q_{\rm tot}}{1 + \epsilon}
\biggr]
\nonumber\\
&=& \frac{F}{2f_a} 
\, \biggl[Q_d - Q_u 
  + \frac{\delta m}{2 \hat{m}} \, (1 - \epsilon)  \, Q_{\rm tot} 
  + {\cal O}(\epsilon^2)\biggr]
\nonumber\\
&=& \frac{F}{2f_a} 
\, \biggl[Q_d - Q_u 
  + \frac{Q_{\rm tot}}{2} \, \frac{\delta m}{\hat{m}}
  \, \Big(1 - \frac{\hat{m}}{2m_s} (1 + 6 x)\Big)
  + {\cal O}\Big(\frac{\hat{m}^2}{m_s^2},
  \frac{\delta m^3}{\hat{m}^3}\Big)\biggr]
\,, \nonumber\\
\stackrel{0}{\theta}_{a\eta} &=&
\frac{F \sqrt{3}}{2f_a} 
 \, \biggl[Q_s + \frac{Q_{GQ}}{3} 
  - \frac{\epsilon \, Q_{\rm tot}}{1 + \epsilon} \, \frac{1 + 2 x}{1 + 6 x} 
\biggr]
\nonumber\\
&=& \frac{F \sqrt{3}}{2f_a} 
 \, \biggl[Q_s + \frac{Q_{GQ}}{3} 
  - \epsilon  \, Q_{\rm tot} \, \frac{1 + 2 x}{1 + 6 x} 
  + {\cal O}(\epsilon^2)\biggr] \,,
\nonumber\\
   \stackrel{0}{\theta}_{a\eta'}
&=& \frac{F \sqrt{3}}{2f_a} 
 \, \biggl[Q_s + \frac{Q_{GQ}}{3} 
  - \frac{Q_{\rm tot}}{2} \, \frac{\hat{m}}{m_s} 
 \, \Big(1 - \frac{\delta m^2}{4 \hat{m}^2}\Big) \, (1 + 2 x)
  + {\cal O}\Big(\frac{\hat{m}^2}{m_s^2}\Big)\biggr]
\,, \nonumber\\
&=& \frac{F}{f_a \sqrt{6}}
\, \biggl[Q_{GQ} - \frac{\epsilon \, Q_{\rm tot}}{1 + \epsilon} 
\, \frac{6 x}{1 + 6 x} 
\biggr]
\nonumber\\
&=& \frac{F}{f_a \sqrt{6}}
\, \biggl[Q_{GQ} - \epsilon \, Q_{\rm tot} \, \frac{6 x}{1 + 6 x} 
  + {\cal O}(\epsilon^2)\biggr]
\nonumber\\
&=& \frac{F}{f_a \sqrt{6}}
\, \biggl[Q_{GQ} - 3 \, Q_{\rm tot} \, \frac{\hat{m}}{m_s} 
\, \Big(1 - \frac{\delta m^2}{4 \hat{m}^2}\Big) \, x
  + {\cal O}\Big(\frac{\hat{m}^2}{m_s^2}\Big)\biggr] \,,
\\[1mm]
M_{a, {\rm QCD}}^2 &=& 
\stackrel{0}{\theta}_{a\pi^0} \Big(\stackrel{0}{\theta}_{a\pi^0}
\, M_{\pi^0}^2 + 2 \, M_{a^0\pi^3} \Big)
\nonumber\\
&+&\stackrel{0}{\theta}_{a\eta} \Big(\stackrel{0}{\theta}_{a\eta}
\, M_{\eta}^2 + 2 \, M_{a^0\eta^8} \Big)
 + \stackrel{0}{\theta}_{a\eta'} \Big(\stackrel{0}{\theta}_{a\eta'}
\, M_{\eta'}^2 + 2 \, M_{a^0\eta^0} \Big)
\nonumber\\
&+& 2 \, \stackrel{0}{\theta}_{a\eta}
    \, \stackrel{0}{\theta}_{a\eta'} \, M_{\eta^8\eta^0}^2 
 +  2 \, \stackrel{0}{\theta}_{a\pi^0}
    \, \Big(\stackrel{0}{\theta}_{a\eta} \, M_{\pi^3\eta^8}
          + \stackrel{0}{\theta}_{a\eta'} \, M_{\pi^3 \eta^0}\Big)  
\nonumber\\
&=& \frac{F^2}{f_a^2} \, \frac{Q_{\rm tot}^2}{1 + \epsilon}
\, \frac{B m_u m_d}{m_u + m_d}
\simeq \frac{F^2}{4 f_a^2} \, Q_{\rm tot}^2 \, M_{\pi}^2 
\nonumber\\
&=& \frac{F^2}{2 f_a^2} \, Q_{\rm tot}^2
\, B \, \hat{m} \, 
\biggl[1 - \frac{\delta m^2}{4 \hat{m}^2} 
  - \frac{\hat{m}}{2m_s}
  \, \Big(1 - \frac{\delta m^2}{2 \hat{m}^2}\Big) \, (1 + 6 x) 
\nonumber\\
&+& {\cal O}\Big(\frac{\hat{m}^2}{m_s^2},\frac{\delta m^4}{\hat{m}^3 m_s}
\Big)\biggr]
\,, 
\en
where $Q_{\rm tot} = Q_u + Q_d + Q_s + Q_{GQ}$,
$\epsilon = \frac{m_u m_d}{m_s (m_u + m_d)} \, (1 + 6 x) 
\sim  \frac{\hat{m}}{m_s} \simeq 0.044$
is the small parameter in which we make an expansion.

\end{document}